%% file: LFVHiggs_paper.tex
\newcommand*{\ATLASLATEXPATH}{latex/}
\documentclass[cernpreprint,texlive=2016,txfonts,UKenglish,biblatex=false]{\ATLASLATEXPATH atlasdoc} 
\pdfoutput=1

\usepackage{\ATLASLATEXPATH atlaspackage}

\usepackage{\ATLASLATEXPATH atlasphysics}

\graphicspath{{logos/}{figures/}{plots/}}

\usepackage{mathrsfs}
\usepackage{cite}
\usepackage{float}
\usepackage{lineno}
\usepackage{multicol}
\usepackage{multirow}
\usepackage{rotating}
\usepackage{morefloats}
\usepackage{rotating}
\usepackage{axodraw}
\usepackage{authblk}
\usepackage{tabulary}
\usepackage{longtable}
\usepackage{ amssymb }
\usepackage{footnote}
\usepackage{lscape}
\usepackage{graphicx} 
\DeclareGraphicsExtensions{.png,.eps}
\usepackage{rotating} 
\usepackage{slashbox} 

\usepackage{LFVHiggs_paper-defs}

\input{LFVHiggs_paper-metadata}

\hypersetup{pdftitle={ATLAS draft},pdfauthor={The ATLAS Collaboration}}

\begin{document}

\maketitle



\input paper_intro
\input paper_objects
\input paper_simulation
\input paper_lephad
\input paper_leplep
\input paper_results
\input Z_selection

\input paper_summary

\section*{Acknowledgements}

\input{Acknowledgements}



\bibliographystyle{bibtex/bst/atlasBibStyleWithTitle}
\bibliography{bibliography,bibtex/bib/ATLAS}
\newpage
\input{atlas_authlist}




\end{document}

%% file: LFVHiggs_paper-metadata.tex
\AtlasTitle{Search for lepton-flavour-violating decays of the Higgs and $Z$ bosons with the ATLAS detector}

\author{The ATLAS Collaboration}

\AtlasRefCode{HIGG-2015-09}

\AtlasJournal{EPJC}
\AtlasJournalRef{Eur. Phys. J. C (2017) 77: 70}
\AtlasDOI{10.1140/epjc/s10052-017-4624-0}

\AtlasAbstract{%
Direct searches for lepton flavour violation in decays of the Higgs
and $Z$ bosons with the ATLAS detector at the LHC 
are presented. The following three decays are considered: $H\to e\tau$, $H\to\mu\tau$, 
and $Z\to\mu\tau$. The searches are based on the data sample of proton--proton 
collisions collected by the ATLAS detector corresponding to an integrated luminosity 
of 20.3~$\ifb$ at a centre-of-mass energy of $\sqrt{s}=8$~\TeV{}. No significant
excess is observed, and upper limits on the lepton-flavour-violating branching ratios are set
at the 95$\%$ confidence level: Br$(H\to e\tau)<\LimObsHe\%$, 
Br$(H\to\mu\tau)<\LimObsHmu\%$, and Br$(Z\to\mu\tau)<\LimObsZmu$.   
}

\AtlasCoverSupportingNote{Higgs LFV LepHad Support Note}{https://cds.cern.ch/record/2033787}
\AtlasCoverSupportingNote{Higgs LFV LepLep Support Note}{https://cds.cern.ch/record/1744055}
\AtlasCoverSupportingNote{Z LFV LepHad Support Note}{https://cds.cern.ch/record/1976182}
\AtlasCoverCommentsDeadline{Mar 6, 2016}

\AtlasCoverAnalysisTeam{R. Clarke, O. Igonkina, A. Pranko, M. Shapiro, H. Weits, L. Fiorini, A. Taffard, N. Ruthmann, D. Gerbaudo, A. Efrati, A. Dery, S. Bressler}

\AtlasCoverEdBoardMember{Erez Etzion, Kevin Black (chair), Lorenzo Bellagamba, Stefania Xella, Swagato Banerjee}

\AtlasCoverEgroupEditors{atlas-higg-2015-09-editors@cern.ch}

\AtlasCoverEgroupEdBoard{atlas-higg-2015-09-editorial-board@cern.ch}

%% file: paper_intro.tex
\section{Introduction}
\label{sec:introduction}

One of the main goals of the Large Hadron Collider (LHC) physics
programme at CERN is to discover physics beyond the Standard Model (SM).
A possible sign would be the observation of lepton flavour violation 
(LFV) that could be realised in decays of the Higgs boson or of the
$Z$ boson to pairs of leptons with different flavours.

Lepton-flavour-violating decays of the Higgs boson can occur naturally in models with more than one Higgs 
doublet~\cite{Bjorken:1977vt,DiazCruz:1999xe,Arana-Catania:2013xma,Arhrib:2012ax}, 
composite Higgs models~\cite{Agashe:2009di,Azatov:2009na}, models with flavour 
symmetries~\cite{Ishimori:2010au}, Randall--Sundrum models~\cite{Perez:2008ee} 
and many others~\cite{Blanke:2008zb,Giudice:2008uua,AguilarSaavedra:2009mx,Albrecht:2009xr,Goudelis:2011un,McKeen:2012av,Crivellin:2015lwa,Crivellin:2015mga}. 
LFV $Z$ boson decays are predicted in models with heavy neutrinos~\cite{Illana:2000ic}, extended gauge models~\cite{Kuo:1985jt} and supersymmetry~\cite{Gabbiani:1988pp}.

The most stringent bounds on the LFV decays of the Higgs and $Z$ bosons other than $H\to{}\mu{}e$ 
are derived from direct searches~\cite{Agashe:2014kda}.
The CMS Collaboration has performed the first direct search for LFV $H\to\mu\tau$ 
decays~\cite{Khachatryan:2015kon} and reported a small excess (2.4 standard deviations) of data 
over the predicted background. Their results give a 1.51$\%$ upper limit on Br($H\to\mu\tau$) at 
the 95$\%$ confidence level (CL). The ATLAS Collaboration has also performed a search~\cite{Aad:2015gha}
for the LFV $H\to\mu\tau$ decays in the final state with one muon and one hadronically decaying $\tau$-lepton,
$\thad{}$, and reported a 1.85$\%$ upper limit on 
Br($H\to\mu\tau$) at the 95$\%$~CL. The most stringent indirect constraint on $H\to e\mu$ decays is 
derived from the results of searches for $\mu\to e\gamma$ decays~\cite{Adam:2013mnn}, and a bound of
Br($H\to e\mu$)~$<$~O($10^{-8}$) is obtained~\cite{Harnik:2012pb,Blankenburg:2012ex}. The bound on 
$\mu\to e\gamma$ decays suggests that the presence of a $H\to\mu\tau$ signal would exclude the 
presence of a $H\to e\tau$ signal, and vice versa, at an experimentally observable level at the 
LHC~\cite{Blankenburg:2012ex}. It is also important to note that a relatively large Br($H\to\mu\tau$) 
can be achieved without any particular tuning of the effective couplings, while a large Br($H\to e\tau$)
is possible only at the cost of some fine-tuning of the corresponding couplings~\cite{Blankenburg:2012ex}.
Upper bounds on the LFV $Z\to e\mu$, $Z\to\mu\tau$ and 
$Z\to e\tau$ decays were set by the LEP experiments~\cite{Akers:1995gz,Abreu:1996mj}: Br$(Z\to e\mu)<1.7\times 10^{-6}$, 
Br$(Z\to e\tau)<9.8\times 10^{-6}$, and Br$(Z\to\mu\tau)<1.2\times 10^{-5}$ at the 95$\%$ CL. 
The ATLAS experiment set the most stringent upper bound on the LFV $Z\to e\mu$ 
decays~\cite{Aad:2014bca}: Br$(Z\to e\mu)<7.5 \times 10^{-7}$ at 95$\%$ CL.

This paper describes three new searches for LFV decays of the Higgs and $Z$ bosons.
The first study is a search for $H\to e\tau$ decays in the final state with one electron and one   
hadronically decaying $\tau$-lepton, $\thad$. The second analysis is a simultaneous search
for the LFV $H\to e\tau$ and $H\to\mu\tau$ decays in the final state
with a leptonically decaying $\tau$-lepton, $\tlep{}$. A combination of
results of the earlier ATLAS search for the LFV $H\to\mu\thad$
decays~\cite{Aad:2015gha} and the two searches described in this paper is also presented. 
The third study constitutes the first ATLAS search for LFV decays
of the $Z$ boson with hadronic $\tau$-lepton decays in the channel $Z\to\mu{}\thad{}$.
The search for LFV decays in the $\tlep{}$ analysis is based on the
novel method introduced in Ref.\cite{Bressler:2014jta}; the searches
in the $\thad{}$ analyses are based on the techniques developed for
the SM $H\to\tlhad$ search.
All three searches are based on the data sample of $pp$ collisions
collected at a centre-of-mass energy of $\sqrt{s}=8$~\TeV{} and
corresponding to an integrated luminosity of 20.3~\ifb.
Given the overlap between the analysis techniques used in
the $H\to{}e\thad$ search and in the $Z\to\mu{}\thad{}$ search, from
here on they are referred to as the $\thad$ channels; the
$H\to{}\ell\tlep$ search is referred to as the $\tlep$ channel, where $\ell{}=e, \mu{}$.




%% file: paper_objects.tex
\section{The ATLAS detector and object reconstruction}
\label{sec:object_reco}

The ATLAS detector\footnote{ATLAS uses a right-handed coordinate system with its origin at 
the nominal interaction point (IP) in the centre of the detector and the $z$-axis along the 
beam pipe. The $x$-axis points from the IP to the centre of the LHC ring, and the $y$-axis 
points upward. Cylindrical coordinates $(r,\phi)$ are used in the transverse plane, $\phi$ 
being the azimuthal angle around the beam pipe. The pseudorapidity is defined in terms of 
the polar angle $\theta$ as $\eta=-\ln\tan(\theta/2)$. The transverse momentum and the 
transverse energy are defined as $\pt=p\times\sin(\theta)$ and $\et=E\times\sin(\theta)$, 
respectively. The distance $\Delta{}R$ in $\eta$--$\phi$ space is defined
as $\Delta$R$~=~\sqrt{(\Delta\eta)^{2} + (\Delta\phi)^{2}}$.} is described in detail in Ref.~\cite{Aad:2008zzm}. ATLAS consists of an inner 
tracking detector (ID) covering the range $|\eta|<2.5$, surrounded by a superconducting 
solenoid providing a 2~T axial magnetic field, a high-granularity
electromagnetic ($|\eta|<3.2$) calorimeter, a hadronic
calorimeter ($|\eta|<4.9$), and a muon spectrometer (MS) ($|\eta|<2.7$) with a toroidal 
magnetic field. 

The signatures of LFV searches reported here are characterised by the presence of an energetic 
lepton originating directly from the boson decay and carrying roughly 
half of its energy, and the hadronic or leptonic decay products of a $\tau$-lepton. The data 
in the $\thad$ channels were collected with single-lepton triggers: a single-muon trigger 
with the threshold of $\pt=24$~\GeV{} and a single-electron trigger with the threshold $\et=24$~\GeV{}.
The data in the $\tlep$ channel were collected using asymmetric electron-muon triggers with  
$(\pt^{\mu},\et^{e})>(18, 8)$~\GeV{} and $(\et^{e},\pt^{\mu})>(14,8)$~\GeV{} thresholds. The 
$\pt$ and $\et$ requirements on the objects in the presented analyses are at least 2~\GeV{} 
higher than the trigger requirements. 

A brief description of the object definitions is provided below. The primary vertex is chosen 
as the proton--proton collision vertex candidate with the highest sum of the squared transverse 
momenta of all associated tracks~\cite{Aad:2015ina}.


Muon candidates are reconstructed using an algorithm that combines 
information from the ID and the MS~\cite{Aad:2014rra}. Muon quality criteria such as inner-detector 
hit requirements are applied to achieve a precise measurement of the muon momentum and 
to reduce the misidentification rate. Muons are required to have $\pt>10$~\GeV{} and to be 
within $|\eta|<2.5$. The distance between the $z$-position of the point of closest approach 
of the muon inner-detector track to the beam-line and the $z$-coordinate of the primary 
vertex is required to be less than 1~cm. In the $\tlep$ channel, there is an additional cut 
on the transverse impact parameter significance, defined as the transverse impact parameter 
divided by its uncertainty: $|d_0|/\sigma_{d_0}<3$. These requirements reduce the contamination 
due to cosmic-ray muons and beam-induced backgrounds. Typical reconstruction and identification 
efficiencies for muons meeting these selection criteria are above 95\%~\cite{Aad:2014rra}.  

Electron candidates are reconstructed from energy clusters in the electromagnetic calorimeters 
matched to tracks in the ID. They are required to have transverse energy $\et{}>15(12)~\GeV{}$
in the \thad{} (\tlep{}) channel,
to be within the pseudorapidity range $|\eta|<2.47$, and to satisfy the 
\textit{medium} shower shape and track selection criteria defined in Ref.~\cite{Aad:2014fxa}. Candidates 
found in the transition region between the barrel and end-cap calorimeters ($1.37<|\eta|<1.52$)
are not considered in the $\thad$ channel. Typical reconstruction and identification efficiencies 
for electrons satisfying these selection criteria range between 80\% and 90\%, depending on $\et$ 
and $\eta$.

Exactly one lepton (electron or muon) satisfying the above identification requirements is 
allowed in the $\thad$ channels. In the $\tlep$ channel, only events with exactly one identified 
muon and one identified electron are retained. All lepton (electron or muon) candidates must
be matched to the corresponding trigger objects and satisfy 
additional isolation criteria, based on tracking and calorimeter information, in order to suppress 
the background from misidentified jets or from semileptonic decays of charm and bottom hadrons.  
The calorimeter isolation variable $I(\et,\Delta R)$ is defined as the sum of the total transverse 
energy in the calorimeter in a cone of size $\Delta{}R$ around the electron cluster or the muon
track, divided by the $\et$ of the electron cluster or the $\pt$ of the muon, respectively. The 
track-based isolation $I(\pt,\Delta R)$ is defined as the scalar sum of the transverse momenta of tracks 
within a cone of size $\Delta R$ around the electron or muon track, divided by the $\et$ of the electron 
cluster or the muon $\pt$, respectively. The contribution due to the lepton itself is not included in either sum. 
The isolation requirements used in the $\thad$ and $\tlep$ 
channels, optimised to reduce the contamination from non-prompt leptons, are listed in Table~\ref{tab:iso}.
\begin{table}[htb]
  \begin{center}
    \caption{Summary of isolation requirements applied for the selection of isolated electrons 
      and muons. The isolation variables are defined in the text.                         
    \label{tab:iso}}
    \begin{tabular}{cll}
      \toprule
                 & $\tlep$ channels        & $\thad$ channels  \\  \midrule
Electrons        & $I(\et,0.3)<0.13$       & $I(\et,0.2)<0.06$ \\
                 & $I(\pt,0.3)<0.07$       & $I(\pt,0.4)<0.06$ \\  \midrule
Muons            & $I(\et,0.3)<0.14$       & $I(\et,0.2)<0.06$ \\
                 & $I(\pt,0.3)<0.06$       & $I(\pt,0.4)<0.06$ \\  \bottomrule
    \end{tabular}
  \end{center}
\end{table}

Hadronically decaying $\tau$-leptons are identified by means of a multivariate analysis technique~\cite{Aad:2014rga}
based on boosted decision trees, which exploits information about ID tracks and clusters in the 
electromagnetic and hadronic calorimeters. The $\thad$ candidates are required to have $+1$ or $-1$ 
net charge in units of electron charge, and must be 1- or 3-track (1- or 3-prong) candidates. 
Events with exactly one $\thad$ candidate satisfying the \textit{medium} identification criteria~\cite{Aad:2014rga} 
with $\pt>20$~\GeV{} and $|\eta|<2.47$ are considered in the $\thad$ channels. In the \tlep{}
channel, events with identified $\thad$ candidates are rejected to avoid overlap between $H\to{}\ell\thad$
and $H\to{}\ell\tlep$. The identification efficiency for $\thad$ candidates satisfying these 
requirements is (55--60)\%. Dedicated criteria~\cite{Aad:2014rga} to separate $\thad$ candidates from 
misidentified electrons are also applied, with a selection efficiency for true $\thad$ decays (that pass 
the $\thad$ identification requirements described above) of 95\%. To reduce the contamination due to 
backgrounds where a muon mimics a $\thad$ signature, events in which an identified muon with $\pt(\mu)>4$~\GeV{} 
overlaps with an identified $\thad$ are rejected~\cite{Aad:2015vsa}. The probability to misidentify a jet 
with $\pt>20$~\GeV{} as a $\thad$ candidate is typically (1--2)\%~\cite{Aad:2014rga}.       

Jets are reconstructed using the anti-$k_{t}$ jet clustering algorithm~\cite{Cacciari:2008gp} with a radius
parameter $R=0.4$, taking the deposited energy in clusters of calorimeter cells as inputs. Fully calibrated 
jets~\cite{Aad:2012vm} are required to be reconstructed in the range $|\eta|<4.5$ and to have $\pt>30$~\GeV{}. 
To suppress jets from multiple proton--proton collisions in the same or nearby beam bunch crossings,
tracking information is used for central jets with $|\eta|<2.4$ and $\pt<50$~\GeV{}.
In the $\tlep$ channel, these central jets are required to have at least one track originating from the primary vertex.
In the $\thad$ channel, tracks originating from the primary vertex must contribute more than half of the jet $\pt$ 
when summing the scalar $\pt$ of all tracks in the jet; jets with no associated tracks are retained.

In the pseudorapidity range $|\eta|<2.5$, jets containing $b$-hadrons ($b$--jets) are selected using a 
tagging algorithm~\cite{ATLAS-CONF-2014-046}. These jets are required
to have $\pt{}>30~\GeV{}$ in the $\thad$ channel, and $\pt{}>20~\GeV{}$
in the $\tlep$ channel.
Two different working points with $\sim$70$\%$ and $\sim$80$\%$ 
$b$-tagging efficiencies for $b$-jets in simulated $t\bar{t}$ events
are used in the $\thad$ and $\tlep$ channels, respectively.
The corresponding light-flavour 
jet misidentification probability is (0.1--1)\%, depending on the $\pt$ and $\eta$ of the jet. Only a very 
small fraction of signal events have $b$--jets, therefore events with identified $b$--jets are vetoed in 
the selection of signal events.

Some objects might be reconstructed as more than one
candidate. Overlapping candidates, separeted by $\Delta R < 0.2$, are
resolved by discarding one object and selecting the other one in the
following order of priority (from highest to lowest): muons,
electrons, $\thad$, and jet candidates~\cite{Aad:2015vsa}.

The missing transverse momentum (with magnitude \met) is reconstructed using the energy deposits in 
calorimeter cells calibrated according to the reconstructed physics objects ($e$, $\gamma$, $\thad$, 
jets and $\mu$) with which they are associated~\cite{Aad:2012re}. In the $\thad$ channels, the 
energy from calorimeter cells not associated with any physics object is included in the $\met$ 
calculation. It is scaled by the scalar sum of $\pt$ of tracks which originate from the 
primary vertex but are not associated with any objects divided by the scalar sum of $\pt$ of all tracks in the 
event which are not associated with objects. The scaling procedure achieves a more accurate reconstruction 
of $\met$ under high pile--up conditions.



%% file: paper_simulation.tex
\section{Signal and background samples}
\label{sec:simulation}

The LFV signal is estimated from simulation. The major Higgs boson production processes (gluon
fusion $ggH$, vector-boson fusion VBF, and associated production $WH/ZH$) are considered in the reported 
searches for LFV $H\to e\tau$ and $H\to \mu\tau$ decays. In the $\tlep$ channel, all backgrounds 
are estimated from data. In the $\thad$ channels, the $Z/\gamma^{*}\to\tau\tau$ and multi-jet backgrounds 
are estimated from data, while the other remaining backgrounds are estimated from simulation, as 
described below.

The largely irreducible $Z/\gamma^{*}\to\tau\tau$ background is modelled by $Z/\gamma^{*}\to\mu\mu$ data events,
where the muon tracks and associated energy deposits in the calorimeters are replaced by the corresponding 
simulated signatures of the final-state particles of the $\tau$-lepton decay. In this approach, essential 
features such as the modelling of the kinematics of the produced boson, the modelling of the hadronic activity 
of the event (jets and underlying event) as well as contributions from pile--up are taken from data. Therefore, 
the dependence on the simulation is minimised and only the $\tau$-lepton decays and the detector response to the 
$\tau$-lepton decay products are based on simulation. This hybrid sample is referred to as embedded data in 
the following. A detailed description of the embedding procedure can
be found in Ref.~\cite{Aad:2015kxa}.

The $W$+jets, $Z/\gamma^{*}\to\mu\mu$ and $Z/\gamma^{*}\to ee$
backgrounds are modelled by the ALPGEN~\cite{Mangano:2002ea} event 
generator interfaced with PYTHIA8~\cite{Sjostrand:2007gs} to provide the parton showering, hadronisation and 
the modelling of the underlying event. The backgrounds with top quarks are modelled by the 
POWHEG~\cite{Nason:2004rx,Frixione:2007vw,Alioli:2010xd} (for $t\bar{t}$, $Wt$ and $s$--channel single-top 
production) and AcerMC~\cite{Kersevan:2004yg} ($t$-channel single-top production) event generators interfaced 
with PYTHIA8. The 
ALPGEN event generator interfaced with HERWIG~\cite{Corcella:2000bw}
is used to model the $WW$ process, and HERWIG is used for the $ZZ$ and
$WZ$ processes.

The events with Higgs bosons produced via $ggH$ or VBF processes are generated at next-to-leading-order (NLO) 
accuracy in QCD with the POWHEG~\cite{Alioli:2008tz} event generator interfaced with PYTHIA8 to provide 
the parton showering, hadronisation and the modelling of the underlying event. The associated production 
($ZH$ and $WH$) samples are simulated using PYTHIA8. All events with Higgs bosons are produced with a mass of 
$m_{H}=125$~\GeV{} assuming the narrow width approximation and normalised to cross sections calculated at 
next-to-next-to-leading order (NNLO) in
QCD~\cite{Anastasiou:2002yz,Ravindran:2003um,Bolzoni:2010xr}. The SM $H\to\tau\tau$ decays are simulated by 
PYTHIA8; the other SM decays of the Higgs boson are negligible.
The LFV Higgs boson decays are modelled by the EvtGen~\cite{Lange:2001uf} event generator according 
to the phase-space model. In the $H\to\mu\tau$ and $H\to e\tau$ decays, the $\tau$-lepton decays are treated 
as unpolarised because the left- and right-handed $\tau$-lepton polarisation states are produced at equal 
rates. Finally, the LFV $Z$ boson decays are simulated with PYTHIA8 assuming an isotropic decay. The width of 
the $Z$ boson is set to its measured value~\cite{Agashe:2014kda}.

For all simulated samples, the decays of $\tau$-leptons are modelled with TAUOLA~\cite{Jadach:1993hs} and 
the propagation of particles through the ATLAS detector is simulated with 
GEANT4~\cite{Aad:2010ah,Agostinelli:2002hh}. The effect of multiple proton--proton collisions in the 
same or nearby beam bunch crossings is accounted for by overlaying additional minimum-bias events.
Simulated events are weighted so that the distribution of the average
number of interactions per bunch crossing matches that observed in
data.

Background contributions due to non-prompt leptons in the $\tlep$ channel and multi-jet events 
in the $\thad$ channel are estimated using data-driven techniques described in Sections~\ref{sec:backgrounds} 
and \ref{sec:ll_bkg}.



%% file: paper_lephad.tex
\section{Search for $H\to e\tau$ decays in the $\thad$ channel}
\label{sec:lephad}

The search for the LFV $H\to e\tau$ decays in the $\thad$ channel
follows exactly the same analysis strategy and utilises the same
background estimation techniques as those used in the
ATLAS search for the LFV $H\to\mu\tau$ decays in the 
$\thad$ channel~\cite{Aad:2015gha}. The only major difference is 
that a high--$\et$ electron is required in the final state instead of
a muon. 
A detailed description of the $H\to e\thad$ analysis is provided in 
the following sections.

\input paper_selection
\input paper_method
\input paper_systematics
\input results_lephad


%% file: paper_selection.tex
\subsection{Event selection and categorisation}
\label{sec:selection}

Signal $H\to e\tau$ events in the $e\thad$ final state are characterised by the presence of exactly one energetic 
electron and one $\thad$ of opposite-sign (OS) charge as well as moderate $\met$, which tends to be aligned with the $\thad$ 
direction.
Same-sign (SS) charge events are used to control the rates of background contributions.
Events with identified muons are rejected.
Backgrounds for this signature can be broadly classified into two major categories:
\begin{itemize}
\item Events with true electron and $\thad$ signatures. These are dominated by the irreducible $Z/\gamma^{*}\to\tau\tau$ production 
with some contributions from the $VV\to e\tau+X$ (where $V=W,Z$), $t\bar{t}$, single-top and SM 
$H\to\tau\tau$ production processes. These events exhibit a very strong charge anti-correlation between 
the electron and the $\thad$. Therefore, the expected number of OS events
($N_{\mathrm{OS}}$) is much larger than the number of SS events ($N_{\mathrm{SS}}$). 
\item Events with a misidentified $\thad$ signature. These are dominated by $W$+jets events with some contribution 
from  multi-jet (many of which have genuine electrons from semileptonic 
decays of heavy-flavour hadrons), diboson ($VV$), $t\bar{t}$ and single-top events  with
$N_{\mathrm{OS}}>N_{\mathrm{SS}}$. Additional contributions to this category arise from $Z(\to ee)$+jets events, 
where a $\thad$ signature can be mimicked by either a jet (no charge correlation) or an electron (strong charge anti-correlation). 
\end{itemize}

\begin{figure}
\centering
\includegraphics[width=0.45\textwidth]{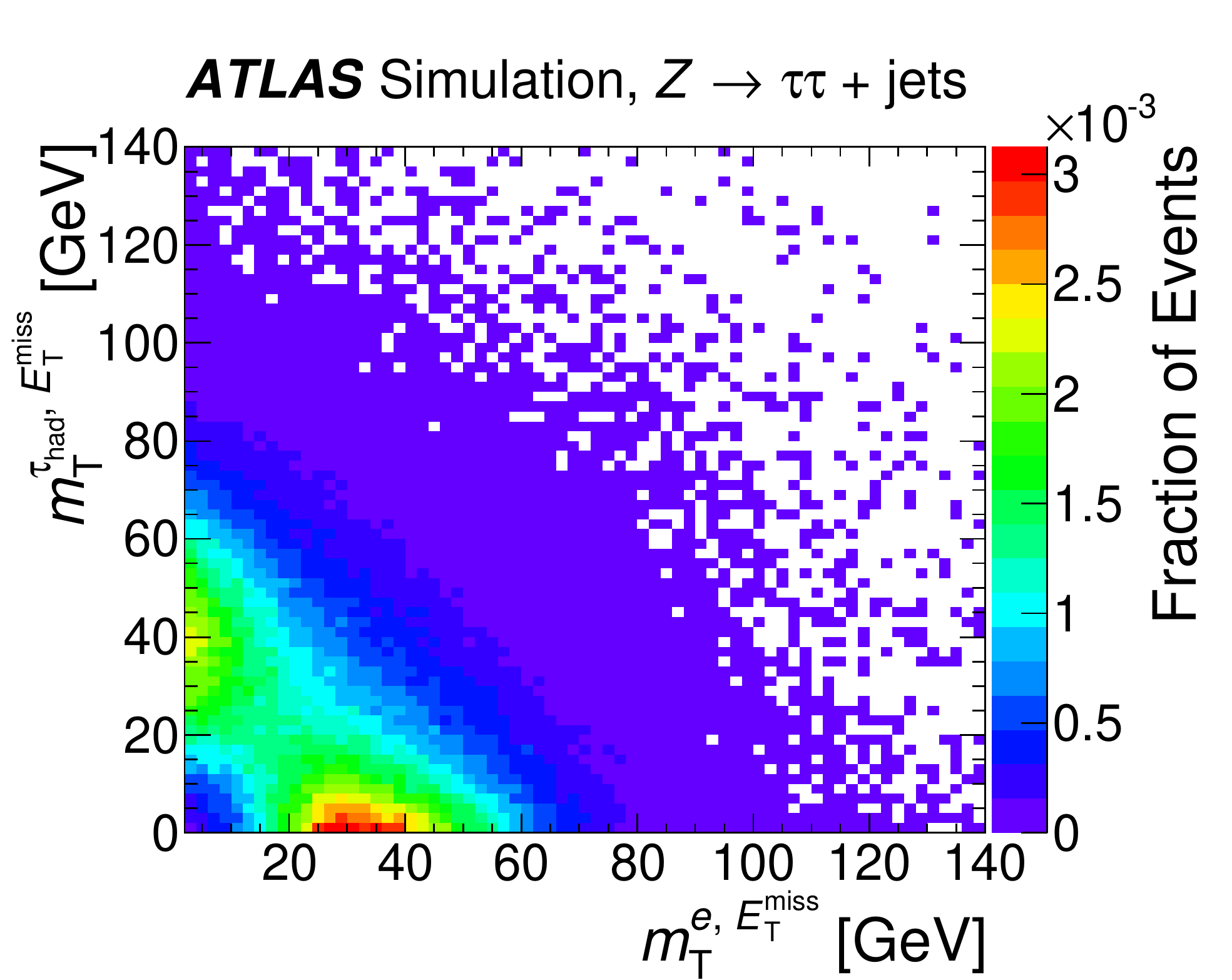}
\includegraphics[width=0.45\textwidth]{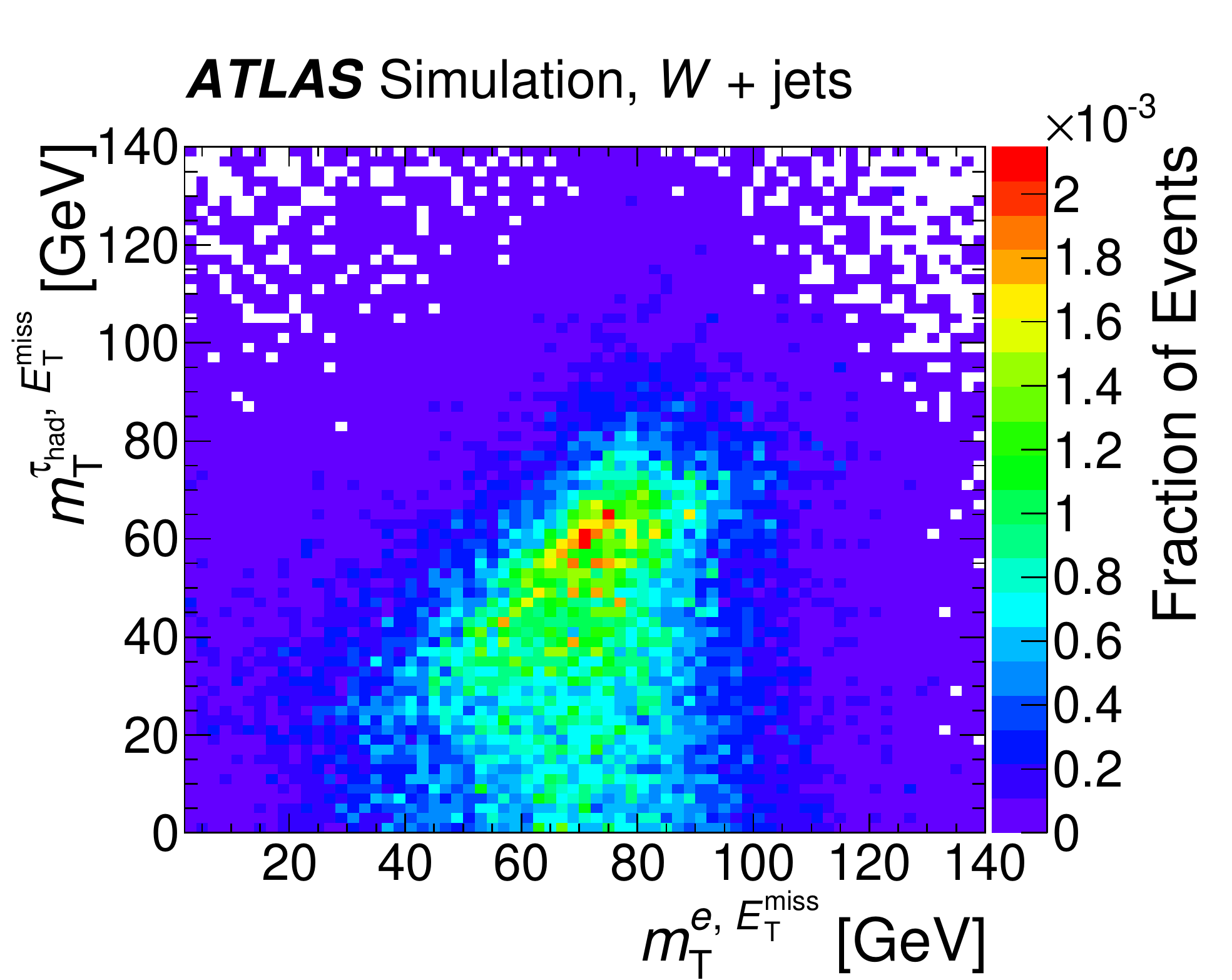}
\includegraphics[width=0.45\textwidth]{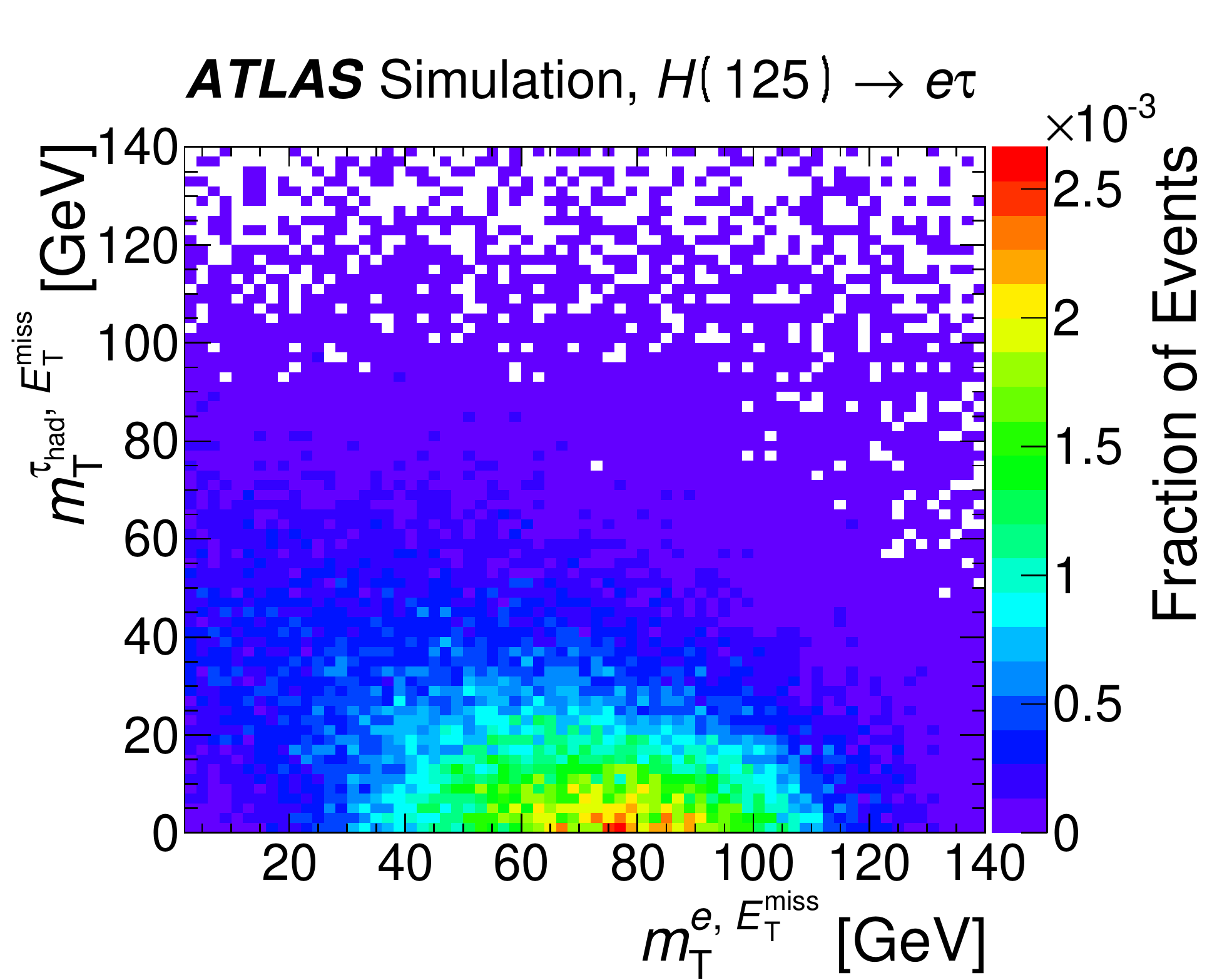}
\includegraphics[width=0.45\textwidth]{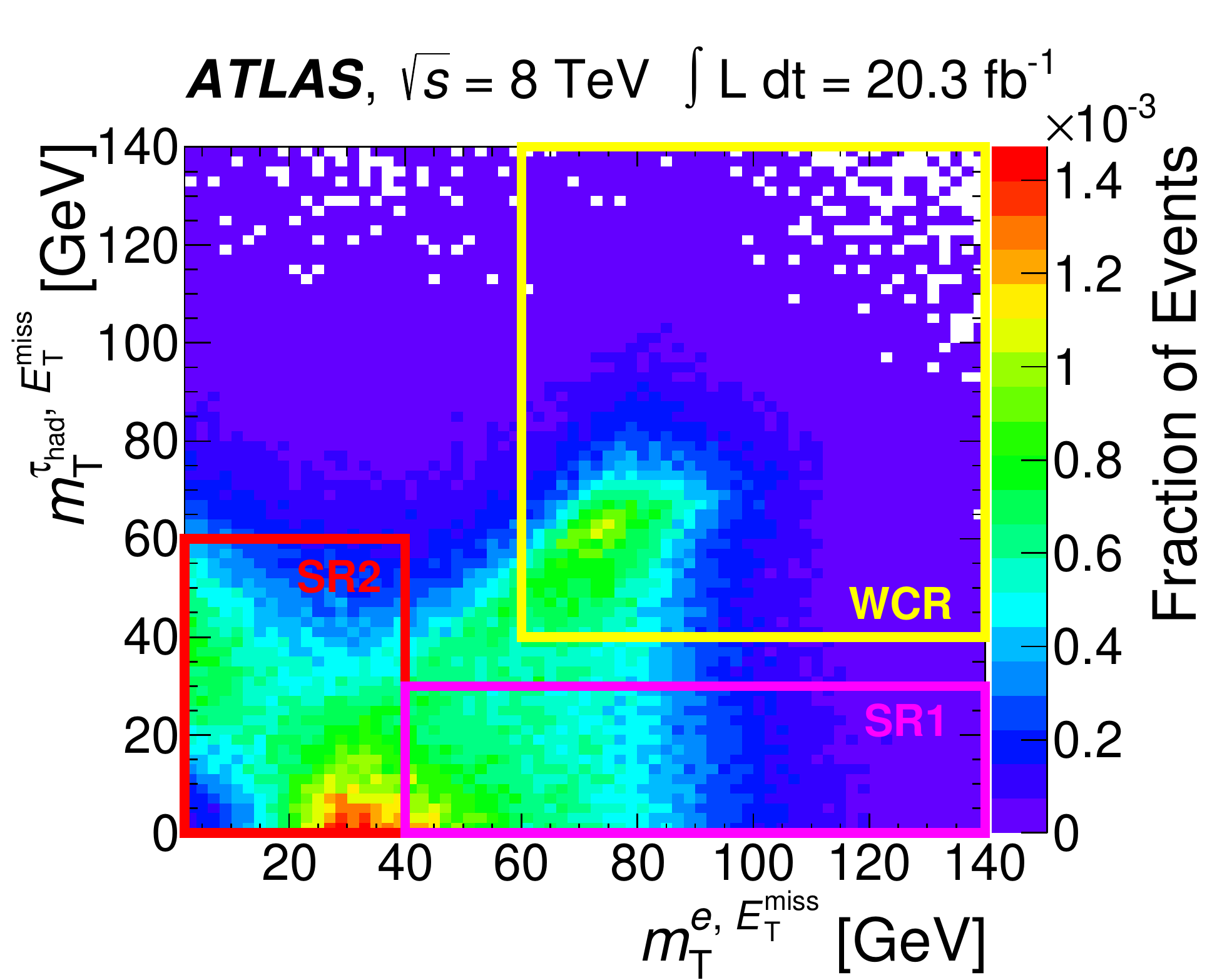}
\caption{Two--dimensional distributions of the transverse mass of the $e$--$\met$ system,
$\mT^{e,\met}$, and that of the $\thad$--$\met$ system, $\mT^{\thad,\met}$, in simulated 
$Z/\gamma^{*}\to\tau\tau$ (top left plot), $W$+jets (top right plot), $H\to e\tau$ signal 
(bottom left plot) and data (bottom right plot) events. Magenta, red and yellow boxes on 
the bottom right plot illustrate SR1, SR2, and WCR, respectively. 
All events used for these distributions are required to have a well-identified electron 
and $\thad$ (as described in text) of opposite charge with $\pt(\thad)>20$~\GeV{} and 
$\et(e)>26$~\GeV{}.}
\label{fig:mt_2Dplots}
\end{figure}

Events with a misidentified $\thad$ tend to have a much softer $\pt(\thad)$ spectrum and a larger angular separation 
between the $\thad$ and $\met$ directions. These properties are exploited to suppress backgrounds and define 
signal and control regions. Events with exactly one electron and exactly one $\thad$ with $\et(e)>26$~\GeV{}, 
$\pt(\thad)>45$~\GeV{} and $|\eta(e)-\eta(\thad)|<2$ form a baseline sample as it represents a common selection for 
both the signal and control regions. The $|\eta(e)-\eta(\thad)|$ cut has $\sim$99\% efficiency for signal and rejects 
a considerable fraction of multi-jet and $W$+jets events. 
Similarly as done in Ref.~\cite{Aad:2015gha}, two signal 
regions are defined using the transverse mass\footnote{$\mT^{\ell,\met}=\sqrt{2\pt^{\ell}\met(1-\cos{\Delta\phi})}$, 
where $\ell=e$,~$\thad$ and $\Delta\phi$ is the azimuthal separation between the directions of the lepton 
($e$ or $\thad$) and $\met$ vectors.}, $\mT$, of the $e$--$\met$ and $\thad$--$\met$ systems: OS events with 
$\mT^{e,\met}>40$~\GeV{} and $\mT^{\thad,\met}<30$~\GeV{} form the signal region--1 (SR1), while OS events 
with $\mT^{e,\met}<40$~\GeV{} and $\mT^{\thad,\met}<60$~\GeV{} form the signal region--2 (SR2). Both
regions have similar sensitivity to the signal (see Section~\ref{sec:lephad_results}). The dominant background
in SR1 is $W$+jets, while the $Z/\gamma^{*}\to\tau\tau$ and $Z\to ee$+jets backgrounds dominate in SR2.   
The modelling of the $W$+jets background is checked in a dedicated control region (WCR) formed by events with 
$\mT^{e,\met}>60$~\GeV{} and $\mT^{\thad,\met}>40$~\GeV{}. As discussed in detail in Section~\ref{sec:backgrounds}, 
the modelling of the $Z/\gamma^{*}\to\tau\tau$ and $Z\to ee$+jets backgrounds is checked in SR2. The choice 
of $\mT$ cuts to define SR1, SR2 and WCR is motivated by correlations between $\mT^{e,\met}$ and $\mT^{\thad,\met}$ 
in $H\to e\tau$ signal and major background ($W$+jets and $Z/\gamma^{*}\to\tau\tau$) events, as illustrated 
in Figure~\ref{fig:mt_2Dplots}. No events with identified $b$--jets are allowed in SR1, SR2 and WCR. The 
modelling of the $t\bar{t}$ and single-top backgrounds is checked in a dedicated control region (TCR), formed 
by events that satisfy the baseline selection and have at least two jets, with at least one being $b$-tagged. 
Table~\ref{tab:cuts} provides a summary of the event selection criteria used to define the signal and control regions. 

The LFV signal is searched for by performing a fit to the mass distribution in data, \Metau, reconstructed 
from the observed electron, $\thad$ and $\met$ objects by means of the Missing Mass Calculator~\cite{mmc} (MMC). 
Conceptually, the MMC is a more sophisticated version of the collinear approximation~\cite{Ellis:1987xu}. 
The main improvement comes from requiring that the relative orientations of the neutrino and other 
$\tau$-lepton decay products are consistent with the mass and kinematics of a $\tau$-lepton decay. 
This is achieved by maximising a probability defined in the kinematically allowed phase-space region.
The MMC used in the $H\to\tau\tau$ analysis~\cite{Aad:2015vsa} is modified to take into account that 
there is only one neutrino from a hadronic $\tau$-lepton decay in LFV $H\to e\tau$ events. For a Higgs 
boson with $m_{H}=125$~\GeV{}, the reconstructed $\Metau$ distribution has a roughly Gaussian shape with a 
full width at half maximum of $\sim$19~\GeV{}. The analysis is performed ``blinded'' in the 
110~\GeV{}$<\Metau<$150~\GeV{} regions of SR1 and SR2, which contain 93.5\% and 95\% of the expected 
signal events in SR1 and SR2, respectively.
The event selection and the analysis strategy are defined without looking at the data in these blinded 
regions. 
 
\begin{table}[tbh]
  \begin{center}
    \caption{Summary of the event selection criteria used to define
      the signal and control regions (see text).                         
    \label{tab:cuts}}
    \begin{tabular}{cllll}
      \toprule
Criterion          & SR1        & SR2        & WCR        & TCR \\  \midrule
$\et(e)$         & $>$26~\GeV{}     & $>$26~\GeV{}     & $>$26~\GeV{}     & $>$26~\GeV{} \\ 
$\pt(\thad)$       & $>$45~\GeV{}     & $>$45~\GeV{}     & $>$45~\GeV{}     & $>$45~\GeV{}  \\
$|\eta(e)-\eta(\thad)|$  & $<$2  & $<$2 & $<$2 & $<$2        \\
$\mT^{e,\met}$     & $>$40~\GeV{}  & $<$40~\GeV{}  & $>$60~\GeV{}  & --  \\
$\mT^{\thad,\met}$   & $<$30~\GeV{}  & $<$60~\GeV{}  & $>$40~\GeV{}  & --   \\
$N_{\mathrm{jet}}$            &  --  &  --  &  --  & $\ge$2    \\
$N_{b-\mathrm{jet}}$          &  0  &  0  &  0  & $\ge$1    \\  \bottomrule
    \end{tabular}
  \end{center}
\end{table}



%% file: paper_method.tex
\subsection{Background estimation}
\label{sec:backgrounds}

The background estimation method takes into account the background properties and composition discussed 
in Section~\ref{sec:selection}. It also relies on the observation that the shape of the $\Metau$
distribution for the multi-jet background is the same for OS and SS events. This observation was made 
using a dedicated control region, MJCR, with an enhanced contribution from the multi-jet background. 
Events in this control region are required to meet all criteria for SR1 and SR2 with the exception of 
the requirement on $|\eta(e)-\eta(\thad)|$, which is reversed: $|\eta(e)-\eta(\thad)|>2$. Therefore, 
the total number of OS background events, $N_{\mathrm{OS}}^{\mathrm{bkg}}$ in each bin of the $\Metau$ (or 
any other) distribution in SR1 and SR2 can be obtained according to the following formula:
\begin{align}
N_{\mathrm{OS}}^{\mathrm{bkg}} =\rqcd \cdot
N_{\mathrm{SS}}^{\mathrm{data}} + \sum_{\mathrm{bkg}-i} N_{\mathrm{OS-SS}}^{\mathrm{bkg}-i},
\label{eq:bckg_formula}
\end{align}
where the individual terms are described below. $N_{\mathrm{SS}}^{\mathrm{data}}$ is the number of SS data events, 
which contains significant contributions from $W$+jets events, multi-jet and other backgrounds. The fractions 
of multi-jet background in SS data events inside the 110~\GeV{}$ < \Metau < $150~\GeV{} mass window are $\sim$27\% 
and $\sim$64\% in SR1 and SR2, respectively. The contributions 
$N_{\mathrm{OS-SS}}^{\mathrm{bkg}-i}=N_{\mathrm{OS}}^{\mathrm{bkg}-i}-\rqcd \cdot N_{\mathrm{SS}}^{\mathrm{bkg}-i}$ are \textit{add-on} 
terms for the different background components (where bkg--$i$ indicates the $i^{\mathrm{th}}$ background source: 
$Z\to\tau\tau$, $Z\to ee$, $W$+jets, $VV$, $H\to\tau\tau$ and events with $t$-quarks), which also account 
for components of these backgrounds already included in SS data events.\footnote{The $\rqcd \cdot N_{\mathrm{SS}}^{\mathrm{bkg}-i}$
  correction in the \textit{add-on} term is needed because same-sign data events include multi-jet as well as 
  electroweak events ($Z\to\tau\tau$, $Z\to ee$, $W$+jets, $VV$, $H\to\tau\tau$ and events with $t$-quarks) 
  and their contributions cannot be separated.} The factor $\rqcd=N_{\mathrm{OS}}^{\mathrm{multi-jet}}/N_{\mathrm{SS}}^{\mathrm{multi-jet}}$ 
accounts for potential differences in flavour composition (and, as a consequence, in jet~$\to\thad$ misidentification rates) 
of final-state jets introduced by the same-sign or opposite-sign charge requirements. The value of $\rqcd=\Rqcd$ 
is obtained from a multi-jet enriched control region in data using a method discussed in Ref.~\cite{HttMoriond2012}.
This sample is obtained by selecting events with $\met<15$~\GeV{}, $\mT^{e,\met}<30$~\GeV{}, removing the isolation criteria 
of the electron candidate and using the \textit{loose} identification criteria for the $\thad$ candidate~\cite{Aad:2014rga}. 
The systematic uncertainty on $\rqcd$ is estimated by varying the selection cuts described above. The obtained value of $\rqcd$ 
is also verified in the MJCR region, which has a smaller number of events but where the electron and $\thad$ candidates pass 
the same identification requirements as events in SR1 and SR2.         

The data and simulation samples used for the modelling of background processes are described in Section~\ref{sec:simulation}. 
A discussion of each background source is provided below.   

The largely irreducible $Z/\gamma^{*}\to\tau\tau$ background is modelled by the embedded data sample described 
in Section~\ref{sec:simulation}. The $Z/\gamma^{*}\to\tau\tau$ normalisation is a free parameter in 
the final fit to data and it is mainly constrained by events with 60~\GeV{}$ < $$\Metau$$ < $90~\GeV{} in SR2.

Events due to the $W$+jets background are mostly selected when the $\thad$ signature is mimicked by jets.
This background is estimated from simulation, and the WCR region is used to check the modelling
of the $W$+jets kinematics
and to obtain separate normalisations for OS and SS $W$+jets events. The difference in these 
two normalisations happens to be statistically significant. An additional overall normalisation 
factor for the $N_{\mathrm{OS-SS}}^{W+\mathrm{jets}}$ term in Eq.~(\ref{eq:bckg_formula}) is introduced as a free 
parameter in the final fit in SR1. By studying WCR events and SR1 events with $\Metau>150$~\GeV{} (dominated by 
$W$+jets background), it is also found that an $\Metau$ shape correction, which depends on the number of jets, 
$\pt(\thad)$ and $|\eta(e)-\eta(\thad)|$, needs to be applied in SR1. This correction is derived from SR1 events with 
$\Metau>150$~\GeV{} and it is applied to events with any value of $\Metau$.  The corresponding modelling uncertainty is 
set to be 50\% of the difference of the $\Metau$ shapes obtained after applying the SR1-based and WCR-based shape corrections.
The size of this uncertainty depends on $\Metau$ and it is as large as $\pm$10\% for 
$W$+jets events with $\Metau < 150$~\GeV{}. In the case of SR2, good modelling of the $N_{\mathrm{jet}}$, $\pt(\thad)$ and
$|\eta(e)-\eta(\thad)|$ distributions suggests that such a correction is not needed. However, a modelling uncertainty 
in the $\Metau$ shape of the $W$+jets background in SR2 is set to be 50\% of the difference between the $\Metau$ 
shape obtained without any correction and the one obtained after applying the correction derived for SR1 events. 
The size of this uncertainty is below 10\% in the 110~\GeV{}$ < \Metau < $150~\GeV{} region, which contains most of 
the signal events. It was also checked that applying the same correction in SR2 as in SR1 would affect the final 
result by less than 4\% (see Section~\ref{sec:results}). 
The modelling of 
jet fragmentation and the underlying event has a significant effect on the estimate of the jet~$\to\thad$ misidentification rate 
in different regions of the phase space and has to be accounted for with a corresponding systematic uncertainty. To 
estimate this effect, the analysis was repeated using a sample of $W$+jets events modelled by ALPGEN interfaced with 
the HERWIG event generator. Differences in the $W$+jets predictions in SR1 and SR2 are found 
to be $\pm$12\% and $\pm$15\%, respectively, and are taken as corresponding systematic uncertainties. 

In the case of the $Z\to ee$ background, there are two components: events in which an electron mimics a $\thad$ 
($e\to\thadmisid{}$) and events in which a jet mimics a $\thad$ (jet$\to\thadmisid{}$). In the first case, 
the shape of the $Z\to ee$ background is obtained from simulation. Corrections from data, derived from 
dedicated tag-and-probe studies~\cite{Aad:2011kt}, are also applied to account for the variation in the 
$e\to\thadmisid{}$ misidentification rate as a function of $\eta$. The normalisation of this background 
component is a free parameter in the final fit to data and it is mainly constrained by events with 
90~\GeV{}$ < $$\Metau$$ < $110~\GeV{} in SR2. For the $Z\to ee$ background where a jet is misidentified 
as a $\thad$ candidate and one of the electrons does not pass the electron identification criteria described
in Section~\ref{sec:object_reco}, the normalisation factor and shape corrections, which depend on the number 
of jets, $\pt(\thad)$ and $|\eta(e)-\eta(\thad)|$, are derived using events with two identified OS electrons 
with an invariant mass, $m_{ee}$, in the range of 80--100~\GeV{}. Since this background does not have an OS--SS charge 
asymmetry, a single correction factor is derived for OS and SS events. Half the difference between the $\Metau$ shape 
with and without this correction is taken as the corresponding systematic uncertainty.

The TCR is used to check the modelling and to obtain normalisations for OS and SS events with top quarks. The normalisation
factors obtained in the TCR are extrapolated into SR1 and SR2, where $t\bar{t}$ and single-top events may have different 
properties. To estimate the uncertainty associated with such an extrapolation, the analysis is repeated using the 
MC@NLO~\cite{Frixione:2002ik} event generator instead of POWHEG for $t\bar{t}$ production.\footnote{The same 
extrapolation uncertainty is assumed for $t\bar{t}$ and single-top backgrounds.} This uncertainty is found to be 
$\pm$8\% ($\pm$14\%) for backgrounds with top quarks in SR1 (SR2).        

The background due to diboson ($WW$, $ZZ$ and $WZ$) production is estimated from simulation, normalised to 
the cross sections calculated at NLO in QCD~\cite{Campbell:2011bn}. Finally, the SM 
$H\to\tau\tau$ events also represent a small background in this search. This background is estimated from 
simulation and normalised to the cross sections calculated at NNLO in
QCD~\cite{Anastasiou:2002yz,Ravindran:2003um,Bolzoni:2010xr}.  All other SM Higgs boson decays constitute negligible 
backgrounds for the LFV signature.

\begin{figure}
\centering
\includegraphics[width=0.45\textwidth]{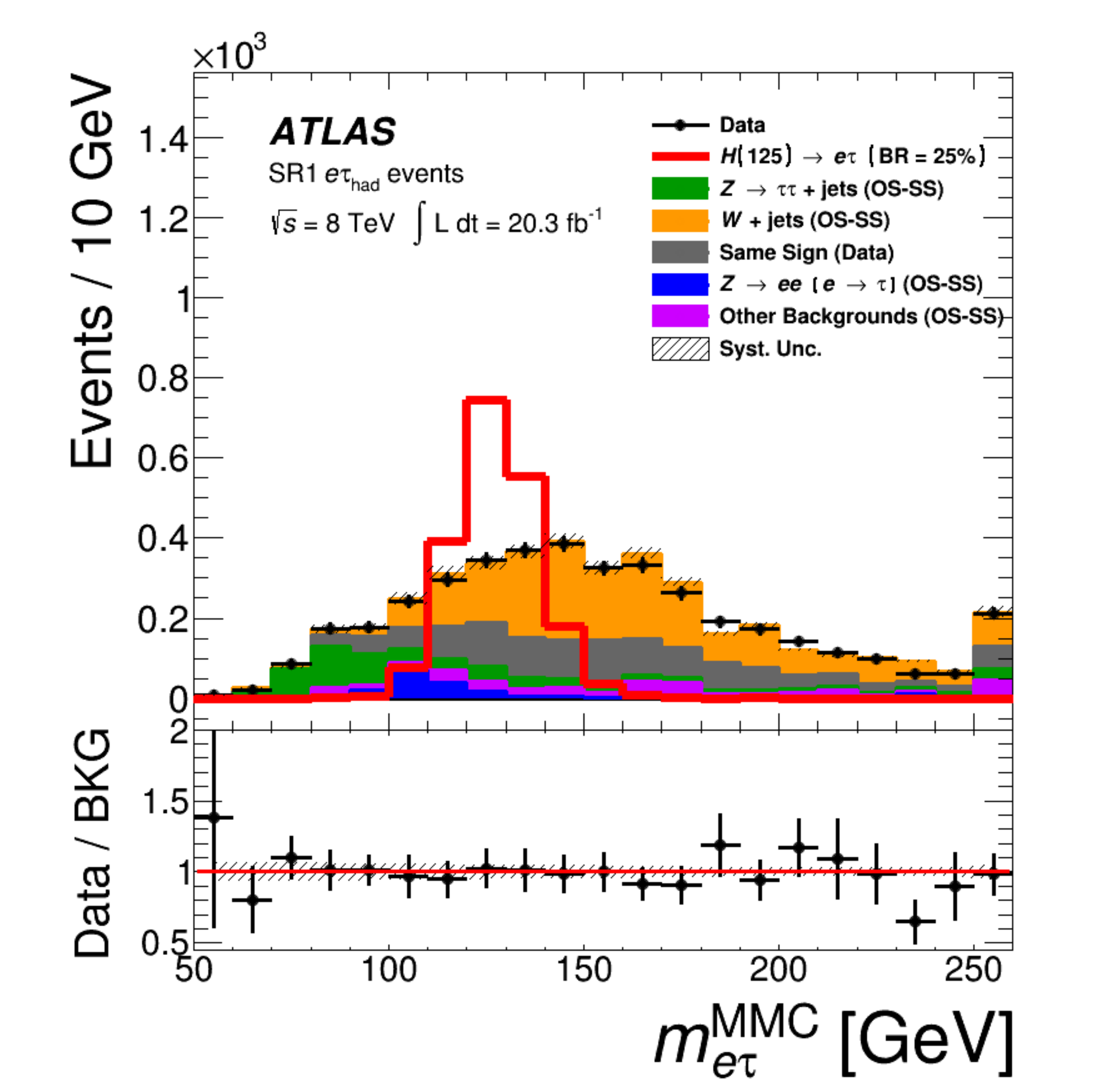}
\includegraphics[width=0.45\textwidth]{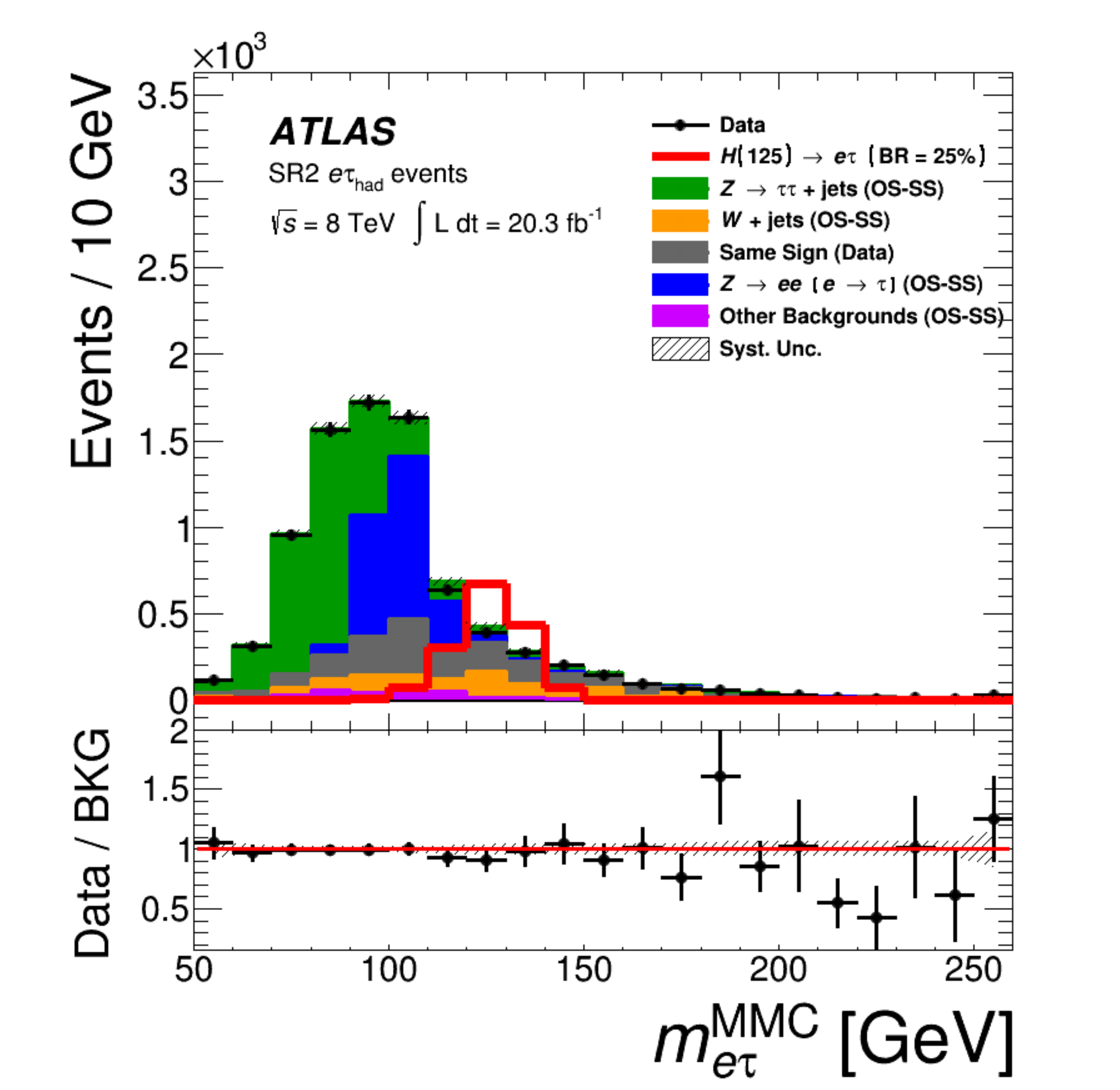}
\caption{Distributions of the mass reconstructed by the Missing Mass
  Calculator, $\Metau$, in SR1 (left) and SR2 (right). 
  The background distributions are determined in a global fit
  (described in Section~\ref{sec:lephad_results}).
  The signal distribution corresponds to Br($H\to e\tau$) = 25\%. The 
  bottom panel of each sub-figure shows the ratio 
  of the observed data to the estimated background. Very small
  backgrounds due to single top, $t\bar{t}$, $VV$, $Z\to ee(\text{jet}\to\thadmisid{})$ and 
  $H\to\tau\tau$ events are combined in a single background component
  labelled as ``Other Backgrounds''. 
  The grey band for the ratio illustrates post-fit systematic 
  uncertainties in the background prediction. The statistical
  uncertainties in the background predictions and data are added 
  in quadrature for the ratios. The last bin in each distribution
  contains events with $\Metau>$~250~\GeV{}.}
\label{fig:postfit_mmc}
\end{figure}

Figure~\ref{fig:postfit_mmc} shows the $\Metau$ distributions for data and the predicted backgrounds in each of 
the signal regions. The backgrounds are estimated using the method described above and their normalisations are 
obtained in a global fit described in Section~\ref{sec:lephad_results}. The signal acceptance times 
efficiencies for passing the SR1 or SR2 selection requirements are 1.8\% and 1.4\%, respectively, and the combined 
efficiency is 3.2\%. The numbers of observed events in the data as well as the signal and background predictions 
in the mass region 110~\GeV{}$ < \Metau < $150~\GeV{} can be found in Table~\ref{tab:yields}.

\begin{table}[htb]
  \begin{center}
    \caption{Data yields, signal and post-fit OS--SS background
      predictions (see Eq.~(\ref{eq:bckg_formula})) for the 
      110~\GeV{}$ < \Metau < $150~\GeV{} region. The signal predictions are 
      given for Br($H\to e\tau$) = 1.0\%. 
      The background predictions are obtained from the combined fit
      to SR1, SR2, WCR and TCR.   
      The post-fit values of systematic uncertainties are 
      provided for the background predictions. 
      For the total background, all correlations between various sources 
      of systematic uncertainties and backgrounds are taken into account.
      The quoted uncertainties represent the statistical (first) and
      systematic (second) uncertainties, respectively.                           
    \label{tab:yields}}
    \begin{tabular}{l r@{}l@{}c@{}r@{}l@{}c@{}r@{}l r@{}l@{}c@{}r@{}l@{}c@{}r@{}l}
      \toprule
                                    & \multicolumn{8}{c}{SR1}                  & \multicolumn{8}{c}{SR2}              \\  \midrule
LFV signal (Br($H\to e\tau$) = 1.0\%) &   75&  &$\pm$&  1&  &$\pm$&   8&   &    59& &$\pm$&  1& &$\pm$&  8& \\ \midrule
$W$+jets                            &  740&    &$\pm$& 80&    &$\pm$& 110&     &   370&   &$\pm$& 60&   &$\pm$& 70&   \\
Same-Sign events                    &  390&    &$\pm$& 20&    &$\pm$&  60&     &   570&   &$\pm$& 30&   &$\pm$& 80&   \\
$Z\to\tau\tau$                      &  116&    &$\pm$&  8&    &$\pm$&  11&     &   245&   &$\pm$& 11&   &$\pm$& 20&   \\
$VV$ and $Z\to ee(jet\to\thadmisid{})$    &   71&    &$\pm$& 31&    &$\pm$&  30&     &    60&   &$\pm$& 20&   &$\pm$& 40&   \\
$Z\to ee(e\to\thadmisid{})$         &   69&    &$\pm$& 17&    &$\pm$&  11&     &   320&   &$\pm$& 40&   &$\pm$& 40&   \\
$t\bar{t}$ and single top                                 &   18&    &$\pm$&  5&    &$\pm$&   4&     &    10&.2 &$\pm$&  2&.6 &$\pm$&  2&.2 \\
$H\to\tau\tau$                      &    4&.6  &$\pm$&  0&.2  &$\pm$&   0&.7   &    10&.5 &$\pm$&  0&.3 &$\pm$&  1&.5 \\
Total background                    & 1410&    &$\pm$& 90&    &$\pm$&  70&     &  1590&   &$\pm$& 80&   &$\pm$& 70&   \\ \midrule
Data                                & 1397&    &     &   &    &     &    &     &  1501&   &     &   &   &     &   &   \\ \bottomrule
    \end{tabular}
  \end{center}
\end{table}



%% file: paper_systematics.tex
\subsection{Systematic uncertainties}
\label{sec:systematics}

The numbers of signal and background events and the shapes of corresponding $\Metau$ distributions are affected by 
systematic uncertainties. They are discussed below and changes in event yields are provided for major sources of 
uncertainties. For all uncertainties, the effects on both the total signal and background predictions and on the 
shape of the $\Metau$ distribution are evaluated. Unless otherwise mentioned, all sources of experimental uncertainties 
are treated as fully correlated across signal and control regions in the final fit which is discussed in 
Section~\ref{sec:lephad_results}.  

The largest systematic uncertainties arise from the normalisation ($\pm$12\% uncertainty) and modelling
of the $W$+jets background. The uncertainties on the $W$+jets normalisation and $\Metau$ shape corrections are treated as 
uncorrelated between SR1 and SR2.  
The uncertainties in $\rqcd$ ($\pm$13\%) and in the 
normalisation ($\pm$13\%) and modelling of $Z\to\tau\tau$ also play an important role. The normalisation 
uncertainty ($\pm$7\%) for the $Z\to ee$ (with $e\to\thadmisid{}$) background has a limited impact on the 
sensitivity because of a good separation of the signal and $Z\to ee$ peaks in the $\Metau$ distribution. The other major 
sources of experimental uncertainty, affecting both the shape and normalisation of signal and backgrounds, are the 
uncertainty in the $\thad$ energy scale~\cite{Aad:2014rga}, which is measured with $\pm$(2--4)\% precision (depending on $\pt$ and decay mode of the $\thad$ candidate), and uncertainties 
in the embedding method used to model the $Z\to\tau\tau$ background~\cite{Aad:2015vsa}. Less significant sources of 
experimental uncertainty, affecting the shape and normalisation of signal and backgrounds, are the uncertainty 
in the jet energy scale~\cite{Aad:2012vm,Aad:2014bia} and resolution~\cite{Aad:2012ag}. The uncertainties in the 
$\thad$ energy resolution, the energy scale and resolution of electrons, and the scale uncertainty in $\MET$ due to 
the energy in calorimeter cells not associated with physics objects are taken into account; however, they are found 
to be only $\pm$(1--2$\%$). The following experimental uncertainties primarily affect the normalisation of signal and 
backgrounds: the $\pm$2.8\% uncertainty in the integrated luminosity~\cite{Aad:2013ucp}, the uncertainty in the $\thad$ 
identification efficiency~\cite{Aad:2014rga}, which is measured to be $\pm$(2--3)\% for 1-prong and $\pm$(3--5)\% 
for 3-prong decays(where the range reflects the dependence on $\pt$ of the $\thad$ candidate), the $\pm$2.1\% uncertainty for triggering, reconstructing and identifying electrons~\cite{Aad:2014fxa}, 
and the $\pm$2\% uncertainty in the $b$-jet tagging efficiency~\cite{ATLAS-CONF-2014-046}.      

Theoretical uncertainties are estimated for the Higgs boson production and for the $VV$ background, which 
are modelled with the simulation and are not normalised to data in dedicated control regions. Uncertainties 
due to missing higher-order QCD corrections in the production cross sections are found to be~\cite{Dittmaier:2011ti} 
$\pm$10.1$\%$ ($\pm$7.8$\%$) for the Higgs boson production via $ggH$ in SR1 (SR2), $\pm$1$\%$ for the $Z\to ee$ 
background and for VBF and $VH$ Higgs boson production, and $\pm$5\% for the $VV$ background. The systematic uncertainties 
due to the choice of parton distribution functions used in the simulation are evaluated based on the prescription described 
in Ref.~\cite{Dittmaier:2011ti} and the following values are used in this analysis: $\pm$7.5$\%$ for the Higgs boson 
production via $ggH$, $\pm$2.8$\%$ for the VBF and $VH$ Higgs boson production, and $\pm$4$\%$ for the $VV$ 
background. Finally, an additional $\pm$5.7\% systematic uncertainty~\cite{Dittmaier:2011ti} on Br($H\to\tau\tau$) is 
applied to the SM $H\to\tau\tau$ background. 



%% file: results_lephad.tex
\subsection{Results of the search for LFV $H\to e\tau$ decays in the $\thad$ channel}
\label{sec:lephad_results}

\begin{figure}
\centering
\includegraphics[width=0.5\textwidth]{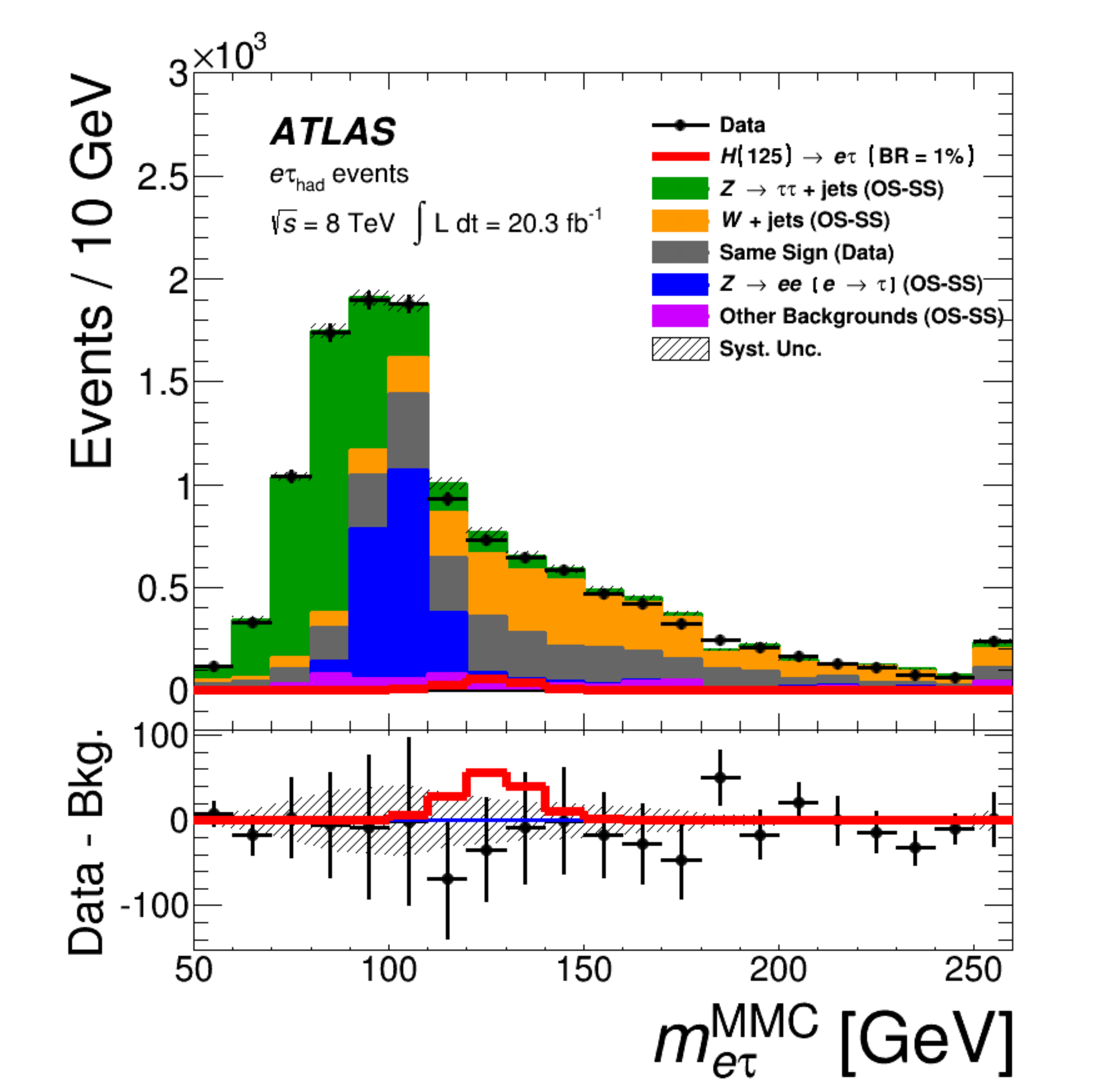}
\caption{Post-fit combined $\Metau$ distribution obtained by adding individual distributions in SR1 
  and SR2. In the lower part of the figure, the data are shown after subtraction of the estimated 
  backgrounds. The grey band in the bottom panel illustrates the post-fit systematic uncertainties in 
  the background prediction. The statistical uncertainties for data and background predictions are added 
  in quadrature in the bottom part of the figure. The signal is shown assuming
  Br$(H\to e\tau)=1.0\%$. Very small backgrounds due to single top, $t\bar{t}$, $VV$, $Z\to ee(jet\to\thadmisid{})$ and 
  $H\to\tau\tau$ events are combined in a single background component labelled as ``Other Backgrounds''. 
  The last bin of the distribution contains events with $\Metau>$250~\GeV{}.}
\label{fig:mmc_combined}
\end{figure}

A simultaneous binned maximum-likelihood fit is performed on the $\Metau$ distributions in SR1 and SR2 and on 
event yields in WCR and TCR to extract the LFV branching ratio Br($H\to e\tau$). The fit exploits the control 
regions and the distinct shapes of the $W$+jets, $Z\to\tau\tau$ and $Z\to ee$ backgrounds in the signal regions 
to constrain some of the systematic uncertainties. This increases the sensitivity of the analysis. The 
post-fit $\Metau$ distributions in SR1 and SR2 are shown in Figure~\ref{fig:postfit_mmc}, and the combined 
$\Metau$ distribution for both signal regions is presented in Figure~\ref{fig:mmc_combined}. Figure~\ref{fig:postfit_mmc} 
illustrates the 
level of agreement between data and background expectations in SR1 and SR2. No statistically significant 
deviations of the data from the predicted background are observed. An upper limit on the LFV branching 
ratio Br($H\to e\tau$) for a Higgs boson with $m_{H}=125$~\GeV{} is set using the CL$_s$ modified frequentist 
formalism~\cite{Read:2002hq} with the test statistic based on the profile likelihood ratio~\cite{Cowan:2010js}. The observed 
and the median expected 95\% CL upper limits are \LimObsHehad\% and \LimExpHehad\%, respectively. Table~\ref{tab:combinedLimits} 
provides a summary of all results, including the results of the ATLAS search for the LFV 
$H\to\mu\tau$ decays~\cite{Aad:2015gha}.



%% file: paper_leplep.tex
\section{Search for $H\to e\tau/\mu\tau$ decays in the $\tlep{}$ channel}
\label{sec:leplep}

In the \tlep{} channel the background estimate is based on the
data-driven method developed in Ref.\cite{Bressler:2014jta}.
This method is sensitive only to the difference between Br($H\to\mu\tau$)
and Br($H\to e\tau$), and it is based on the premise that the
kinematic properties of the SM background are to a good approximation
symmetric under the exchange $e\leftrightarrow{}\mu$.

\input ll_selection
\input ll_bkg
\input ll_systematics

\input ll_stat
\input ll_validation
\input results_leplep

%% file: ll_selection.tex
\subsection{Event selection and signal region definition}
\label{sec:leplepSelection}

Events selected in the $\tlep$ channel must contain exactly two opposite-sign
leptons, one an electron and the other a muon. The lepton with the higher \pt{}
is indicated by $\ell_1$ and the other by $\ell_2$.
Additional kinematic criteria, based on the \pt{} difference between
the two leptons and on the angular separations between the leptons and
the missing transverse momentum, are applied to suppress the SM background
events, which are mainly due to the production of
$Z/\gamma^{*}\to\tau\tau$ and of diboson $(VV)$ events. Two mutually exclusive
signal regions are defined: one with no central ($|\eta|<2.4$) light-flavour jets,
$\mathrm{SR_{noJets}}$, and the other with one or more central
light-flavoured jets, $\mathrm{SR_{withJets}}$. The kinematic criteria
defining each signal region, summarised in Table~\ref{tab:SRDefinition}, are
optimised following two guidelines. The first one is to maximise the
signal-to-background ratio.
The second one is to have, in each signal region, enough
events to perform the data-driven background estimation described
in Section~\ref{sec:ll_bkg}.
\begin{table}[!!!!h]
\begin{center}
\caption{Summary of the selection criteria used to define the signal regions in the \tlep{} channel (see text).}
\label{tab:SRDefinition}
\begin{tabular}{lcc} 
\toprule
  & \SRnoJets & \SRwithJets \\
\midrule
Light leptons &   $e^{\pm}\mu^{\mp}$  & $e^{\pm}\mu^{\mp}$ \\
$\thad$ leptons &  veto   &  veto \\
Central jets & 0 & $\geq 1$ \\
$b$-jets & 0 & 0 \\
$\pt^{\ell_1}$ &  $\geq 35 \GeV$   & $\geq 35 \GeV$ \\
$\pt^{\ell_2}$ &  $\geq 12 \GeV$   & $\geq 12 \GeV$ \\
$\left|\eta^{e}\right|$ &  $\leq 2.4$   & $\leq 2.4$ \\
$\left|\eta^{\mu}\right|$ &  $\leq 2.4$   & $\leq 2.4$ \\
$\Delta \phi (\ell_2,\met)$ &  $\leq 0.7$   & $\leq 0.5$ \\
$\Delta \phi (\ell_1,\ell_2)$ &  $\geq 2.3$   & $\geq 1.0$ \\
$\Delta \phi (\ell_1,\met) $ &  $\geq 2.5$   & $\geq 1.0$ \\
$\Delta\pt(\ell_1,\ell_2)$ &  $\geq 7 \GeV$   & $\geq 1 \GeV$ \\
\bottomrule
\end{tabular}
\end{center}
\end{table}

The final discriminant used in the $\tlep$ channel is the collinear
mass $\mathrm{m_{coll}}$ defined as:
\begin{equation}
  \mcol{} = \sqrt{2\ptZero{}
    \left(\ptOne{}+\met{}\right)
    \left(\cosh{}\Delta\eta-\cos\Delta\phi\right)}.
\end{equation}
This quantity is the invariant mass of two massless particles,
$\tau{}$ and $\ell_1$, computed with the approximation that the decay
products of the $\tau{}$ lepton, $\ell_2$ and neutrinos, are collinear to the
$\tau{}$, and that the \met{} originates from the $\nu{}$.  In the
\htm~(\hte) decay, $\ell_1$~is the muon (electron) and $\ell_2$~is the
electron (muon). The differences in rapidity and azimuthal angle
between $\ell_1$~and $\ell_2$ are indicated by $\Delta\eta{}$ and
$\Delta\phi{}$.
More sophisticated kinematic variables, such as MMC, do not
significantly improve the sensitivity of the \tlep{} channel.



%% file: ll_bkg.tex
\subsection{Background estimation}
\label{sec:ll_bkg}

For simplicity, the symmetry method is illustrated here 
assuming a \htm{}~signal. The same procedure, but with $e$
and $\mu{}$ exchanged, is valid under the \hte{} assumption.
The symmetry method is based on the following two premises:
\begin{enumerate}
\item SM processes result in data that are symmetric
  under the exchange of prompt electrons with prompt muons to a good approximation. In other words,
  the kinematic distributions of prompt electrons and prompt muons are
  approximately the same;\footnote{The effect of the mass difference
    between electrons and muons is negligible for the processes involved. 
  }

\item flavour-violating decays of the Higgs boson break this symmetry.
\end{enumerate}
Dilepton events in the dataset are divided into two mutually
exclusive samples:
\begin{itemize}
\item $\boldsymbol{\mu{}e}$ \textbf{sample}: $\ell_1$~is the
  muon and $\ell_2$~is the electron ($\pt{}^\mu \geq \pt{}^e$)
\item $\boldsymbol{e\mu{}}$ \textbf{sample}: $\ell_1$~is the
  electron and $\ell_2$~is the muon ($\pt{}^e > \pt{}^\mu$)
\end{itemize}

With these assumptions, the SM background is split equally between
the two samples. The \htm{}~signal, however, is present only in the \mes{}
sample because the \pt{} spectrum of electrons from \htm{} decays is
softer then the muon \pt{} spectrum. The number of \htm{} events
in the \ems{} sample is negligible with the selection criteria
described in Section~\ref{sec:leplepSelection}.

For SM events the distributions of kinematic variables
in the two samples are the same with good approximation.  In particular, the collinear mass
distribution differs between the two samples only for the narrow
signal peak. The peak, present only in the distribution of the \mes{}
sample, is on top of the SM background, which, to a good approximation,
can be modelled from the \ems{} collinear mass distribution.

\subsubsection{Asymmetries in the SM background}
\label{sec:asymmetricBkg}

Although the \ems{}--\mes{} symmetry hypothesis is a good starting
assumption, there are effects that can invalidate it and that need to be accounted for. The
first effect is due to events containing misidentified and non-prompt
leptons, together referred to as \textit{non-prompt} in the following.
These leptons originate from misidentified jets or from hadronic decays within
jets. They contribute differently to the \mes{} and \ems{} samples
because the origin of the non-prompt lepton is
different for electrons and for muons. The second effect originates from the
different dependencies on \pt{} and $|\eta|$ that the trigger
efficiency and reconstruction efficiency can have for electrons and
muons. The non-prompt effect is accounted for by estimating the
non-prompt background separately from the other backgrounds. The
efficiency effect is accounted for by scaling the \mcol{} distribution
of the \ems{} sample with a scale factor parameterised as a function of
the sub-leading lepton \pt{}, \ptOne{}.
As shown in Section~\ref{sec:leplepValidation}, the \ems{}-\mes{}
symmetry is restored when these two effects are taken into
account. Smaller effects, which might depend on other parameters such
as $\eta$ or \ptZero{}, are found to be negligible.

\paragraph{Events containing non-prompt leptons}
The background contribution due to non-prompt leptons is estimated
with the matrix method described in Refs.~\cite{ATLAS:2L8TeV,Behnke:2013pga},
which relies on the difference in identification efficiency
between prompt and non-prompt leptons. Two lepton categories are defined:
tight leptons, which must satisfy all the lepton identification
criteria described in Section~\ref{sec:object_reco}, and loose leptons,
which are not required to satisfy the primary vertex and isolation
criteria. By measuring separately for prompt and non-prompt leptons
the tight-to-loose lepton efficiencies, defined as the fraction of
loose leptons that are also tight, one can determine the non-prompt
background contribution from the number of data events that have two
leptons that are either loose or tight. The efficiencies
for prompt and non-prompt leptons, parameterised as a function of \pt{}
and $\eta$, are derived from data with the tag-and-probe
method. Prompt efficiencies are derived from an opposite-sign sample
enriched in $Z\to{}e^{\pm}e^{\mp}$ and
$Z\to{}\mu^{\pm}\mu^{\mp}$. Non-prompt efficiencies are derived from a
same-sign sample ($\mu^{\pm}e^{\pm}$ or $\mu^{\pm}\mu^{\pm}$) where
the muon is the tag lepton.

\paragraph{Asymmetry induced by the different trigger and reconstruction efficiency of electrons and muons}
The efficiency to trigger on and reconstruct an \ems{} event,
$\varepsilon^{e\mu}$, is different from the one of a \mes{} event,
$\varepsilon^{\mu{}e}$. These two efficiencies can be expressed as a
function of the \pt{} of the two leptons:
\begin{equation*}
  \varepsilon^{\mu e}=
  \varepsilon^{\mu e}_{\text{trig.}}\left(\pt^{\ell_{2}=e}\right)\times
  \varepsilon^{\mu}_{\text{reco.}}\left(\pt^{\ell_1=\mu}\right)\times
  \varepsilon^{e}_{\text{reco.}}\left(\pt^{\ell_2=e}\right)
\end{equation*}
\begin{equation*}
  \varepsilon^{e \mu}=
  \varepsilon^{e \mu}_{\text{trig.}}\left(\pt^{\ell_{2}=\mu}\right)\times
  \varepsilon^{e}_{\text{reco.}}\left(\pt^{\ell_1=e}\right)\times
  \varepsilon^{\mu}_{\text{reco.}}\left(\pt^{\ell_2=\mu}\right).
\end{equation*}
%
%
In this search, the leading lepton is required to have
$\ptZero{}>35~\GeV{}$, which is on the plateau region of
the trigger and reconstruction efficiencies. Hence the ratio of
the efficiencies can be approximated as:
\begin{align*}
  \frac{\varepsilon^{\mu e}}{\varepsilon^{e \mu}}
  & =
  \frac{
    \varepsilon^{\mu e}_{\text{trig.}}\left(\pt^{\ell_{2}}\right)
    \varepsilon^{\mu}_{\text{reco.}}\left(\pt^{\ell_1}\right)
    \varepsilon^{e}_{\text{reco.}}\left(\pt^{\ell_2}\right)
  }{
    \varepsilon^{e\mu}_{\text{trig.}}\left(\pt^{\ell_{2}}\right)
    \varepsilon^{e}_{\text{reco.}}\left(\pt^{\ell_1}\right)
    \varepsilon^{\mu}_{\text{reco.}}\left(\pt^{\ell_2}\right)
  } \\
  & =
  \frac{
    \varepsilon^{\mu e}_{\text{trig.}}\left(\pt^{\ell_{2}}\right)
    \varepsilon^{e}_{\text{reco.}}\left(\pt^{\ell_2}\right)
  }{
    \varepsilon^{e\mu}_{\text{trig.}}\left(\pt^{\ell_{2}}\right)
    \varepsilon^{\mu}_{\text{reco.}}\left(\pt^{\ell_2}\right)
  }\times
  \frac{
    \varepsilon^{\mu}_{\text{reco.}}\left(\pt^{\ell_1}\right)
  }{
    \varepsilon^{e}_{\text{reco.}}\left(\pt^{\ell_1}\right)
  } \\
  & =
  \fptOne{}\times\text{Const.}
\end{align*}

Therefore, the ratio of the \ems{} and \mes{} event reconstruction
efficiencies can be parameterised as a function of the sub-leading
lepton \pt{}, \fptOne{}.
Using the fit described in Section~\ref{sec:leplepStat}, the parameter
\fptOne{} is determined in three \ptOne{} bins, 12--20~\GeV{},
20--30~\GeV{}, and $> 30~\GeV{}$.



%% file: ll_systematics.tex
\subsection{Systematic uncertainties}
\label{sec:leplepSystematics}

Using the $e\mu{}$ asymmetry technique, the only systematic uncertainty
associated with the background prediction is due to the non-prompt
background modelling. This uncertainty has two components: the first
one is the limited number of tag-and-probe events used to extract the
prompt and non-prompt efficiencies; the second one is the difference
in kinematics, and therefore in sources of non-prompt leptons, between
the events used to extract the non-prompt efficiency and the events in
the signal regions. This second component is evaluated by measuring
the non-prompt efficiencies in subsets of the nominal tag-and-probe
sample. The subsets are obtained by applying, one at a time, the
kinematic requirements of the signal regions. The ensuing
uncertainties in the estimated number of non-prompt events can be as
large as 10--50$\%$ for the non-prompt efficiency and $3\%$ for the
prompt efficiency, depending on the signal region.

Uncertainties related to the signal prediction are the same ones
described in Section~\ref{sec:systematics} with one minor difference in
the uncertainty in the signal cross section due to higher-order QCD
corrections. This uncertainty is split into two anticorrelated
components: $\pm{}12\%$ in \SRwithJets{} and $\pm{}20\%$ in
\SRnoJets{}.


%% file: ll_stat.tex
\subsection{The statistical model}
\label{sec:leplepStat}

Assuming that the SM background is completely symmetric
when exchanging $e\leftrightarrow\mu{}$ , the likelihood function
for the collinear mass distribution of the \ems{} and \mes{} samples
can be written as:
\begin{equation}
L(b_i,\mu) =
      \prod\limits_i^{N_{\mcol{}}} \mathrm{Pois}(n_i\mid b_i) \times \mathrm{Pois}(m_i\mid b_i+\mu s_i),
\label{equ:coreLikelihood}
\end{equation}
where $n_i$ $(m_i)$ is the number of \ems{} (\mes{}) events in the
$i$-th of the $N_{\mcol}$ \mcol{} bins. The number of background
events in the $i$-th \mcol{} bin is indicated by $b_i$, and $s_i$ is
the number of \htm {} events in the $i$-th mass bin.
The number of signal events $\sum\limits_{i}s_i$ is
normalised to a branching ratio $\mbox{Br}(\htm{})=1\%$, multiplied by
a signal strength $\mu$. The likelihood for the \mcol{} distributions
with a \hte{} signal can be defined in a similar way.  The
contributions due to non-prompt leptons add to the \ems{} and \mes{}
terms and they are denoted by $N_i^{\text{np}}$ and $M_i^{\text{np}}$, along
with their uncertainties, $\sigma_{N_i^{\text{np}}}$ and
$\sigma_{M_i^{\text{np}}}$.
The numbers of non-prompt events in each bin, $N_i^{\text{np}}$ and
$M_i^{\text{np}}$, are treated as Gaussian nuisance parameters.

The \fptOne{} correction, described in Section~\ref{sec:ll_bkg}, is
implemented by performing the fit separately in $N_{\ptOne{}}=3$ \ptOne{} bins,
labelled with the index $j$. The corrective scale factor $A_j$,
corresponding to the \fptOne{} value in the \mcol{} bin $i$ and
\ptOne{} bin $j$, multiplies the \ems{} yield $b_{ij}$. These scale
factors are treated in the statistical model as unconstrained nuisance parameters.

Adding up the symmetric contribution ($b_{ij}$), the non-prompt
contributions ($N_{ij}^{\text{np}}$ and $M_{ij}^{\text{np}}$), the $\fptOne{}$
correction, and the signal contribution ($s_{ij}$), the likelihood is
written as:
%
\begin{equation}
\label{equ:turnOnCurves}
\begin{split}
  L(\mu, b_{ij}, n_{ij}^{\text{np}}, m_{ij}^{\text{np}}) & =
  \prod\limits_i^{N_{\mcol{}}} \prod\limits_j^{N_{\ptOne{}}}
  \mathrm{Pois}(n_{ij}\mid A_jb_{ij} + n_{ij}^{\text{np}}) \times
  \mathrm{Pois}(m_{ij}\mid b_{ij} + m_{ij}^{\text{np}} + \mu s_{ij}) \\ & \times
  \mathrm{Gaus}(n_{ij}^{\text{np}} | N_{ij}^{\text{np}} , \sigma_{N_{ij}^{\text{np}}}) \times
  \mathrm{Gaus}(m_{ij}^{\text{np}} | M_{ij}^{\text{np}} , \sigma_{M_{ij}^{\text{np}}}).
\end{split}    
\end{equation}


%% file: ll_validation.tex
\subsection{Background model validation}
\label{sec:leplepValidation}


The symmetry-based method is validated with simulation and with
data. The validation with simulated samples is performed by comparing
the signal strength measured in the SR with background samples, and
with signal samples corresponding to several non-zero LFV branching
ratios. The validation with data is performed in a validation region
(VR) defined as \SRnoJets{}, but with at least one angular requirement
reversed, $\Delta\phi(\ell_1,\ell_2)$ or $\Delta\phi(\ell_1,\met)$.

The validation procedure consists of comparing the data, or the sum of
the simulated background samples, to the total background
estimated from the statistical model. The comparison is done  
for the \ems{} sample and the \mes{} one. With the simulated
samples, it is also verified that the symmetric background and the
\fptOne{} do not depend on the presence of an LFV signal. 

Generated pseudo-experiments are used to confirm that the statistical
model is unbiased. No significant discrepancy was found between the
injected signal strength and its fitted value up to LFV branching ratios of $10\%$.


%% file: results_leplep.tex
\subsection{Results of the search for LFV $H\to e\tau/\mu\tau$ decays in the $\tlep{}$ channel}
\label{sec:leplep_results}

\begin{figure}[htb]
\begin{center}
\includegraphics[width=0.475\textwidth]{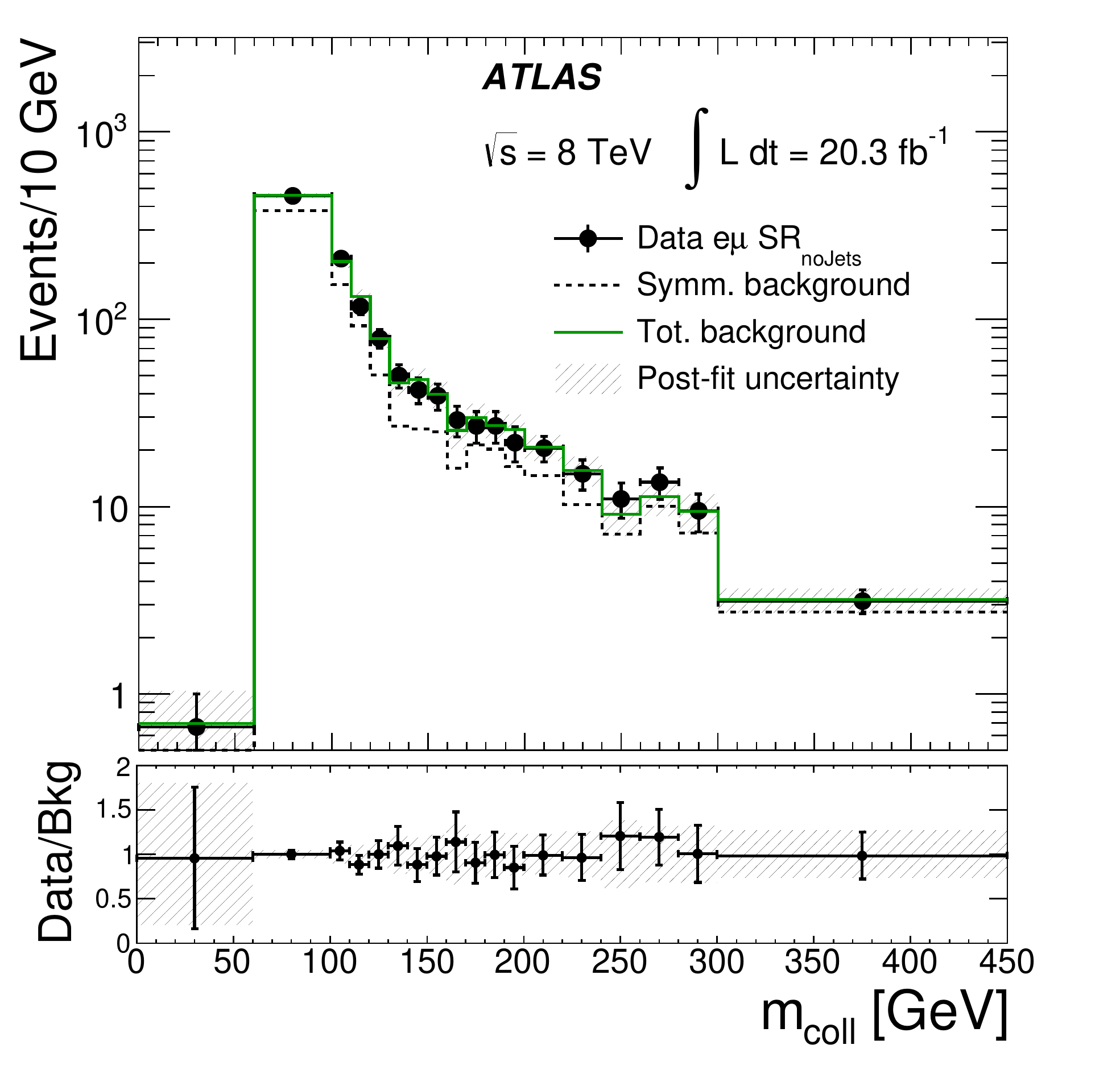}
\includegraphics[width=0.475\textwidth]{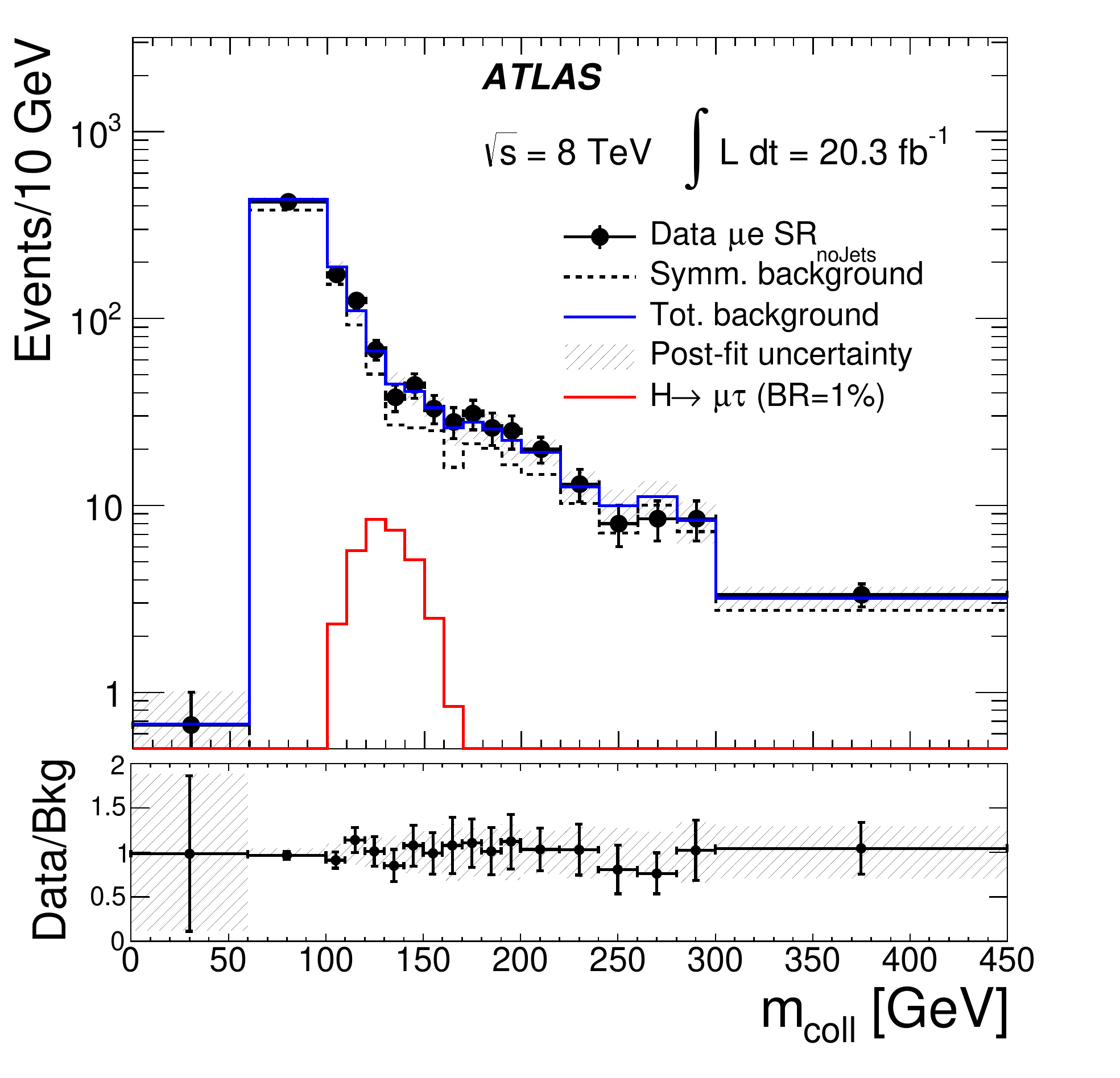}
\\
\includegraphics[width=0.475\textwidth]{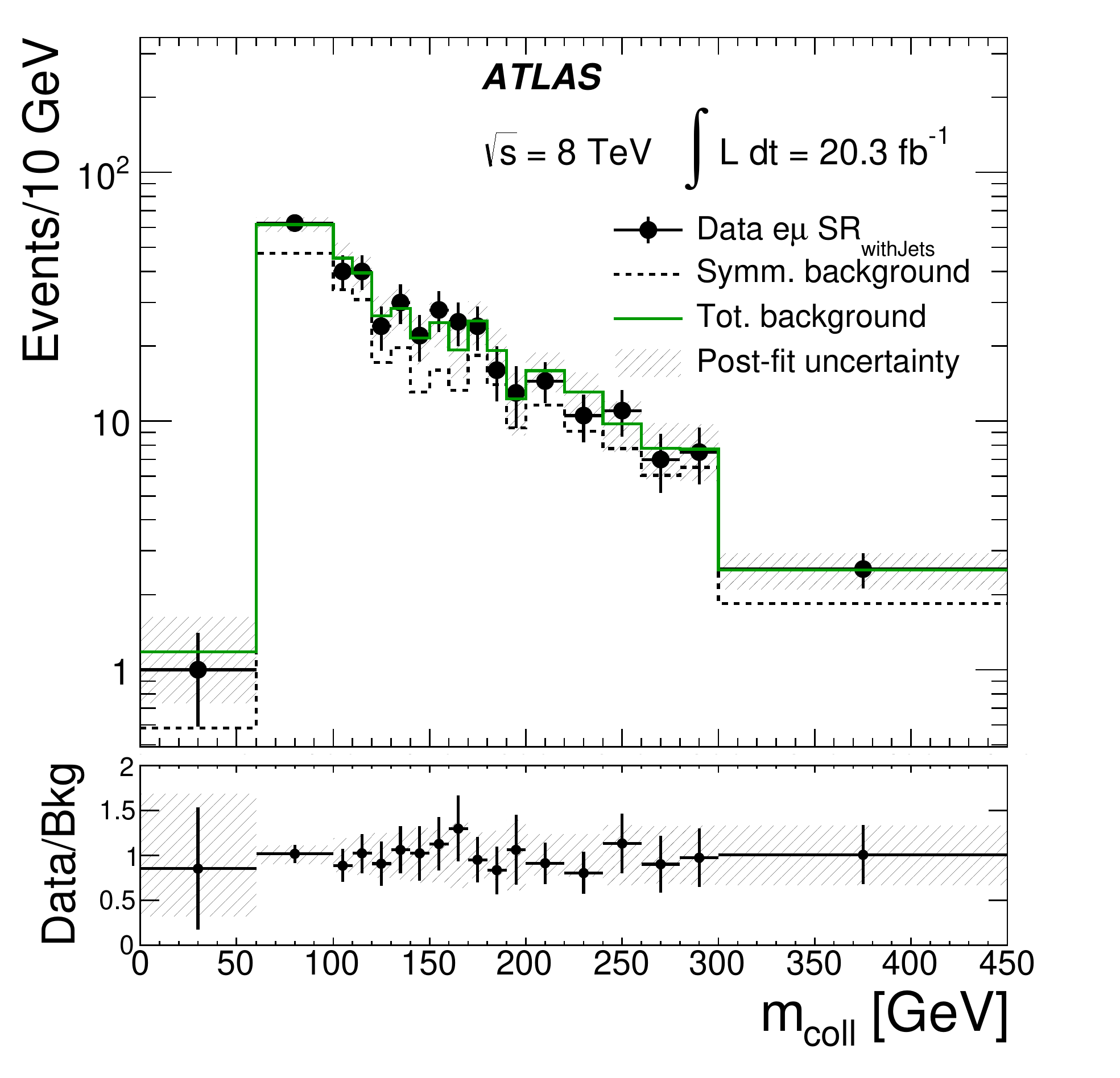}
\includegraphics[width=0.475\textwidth]{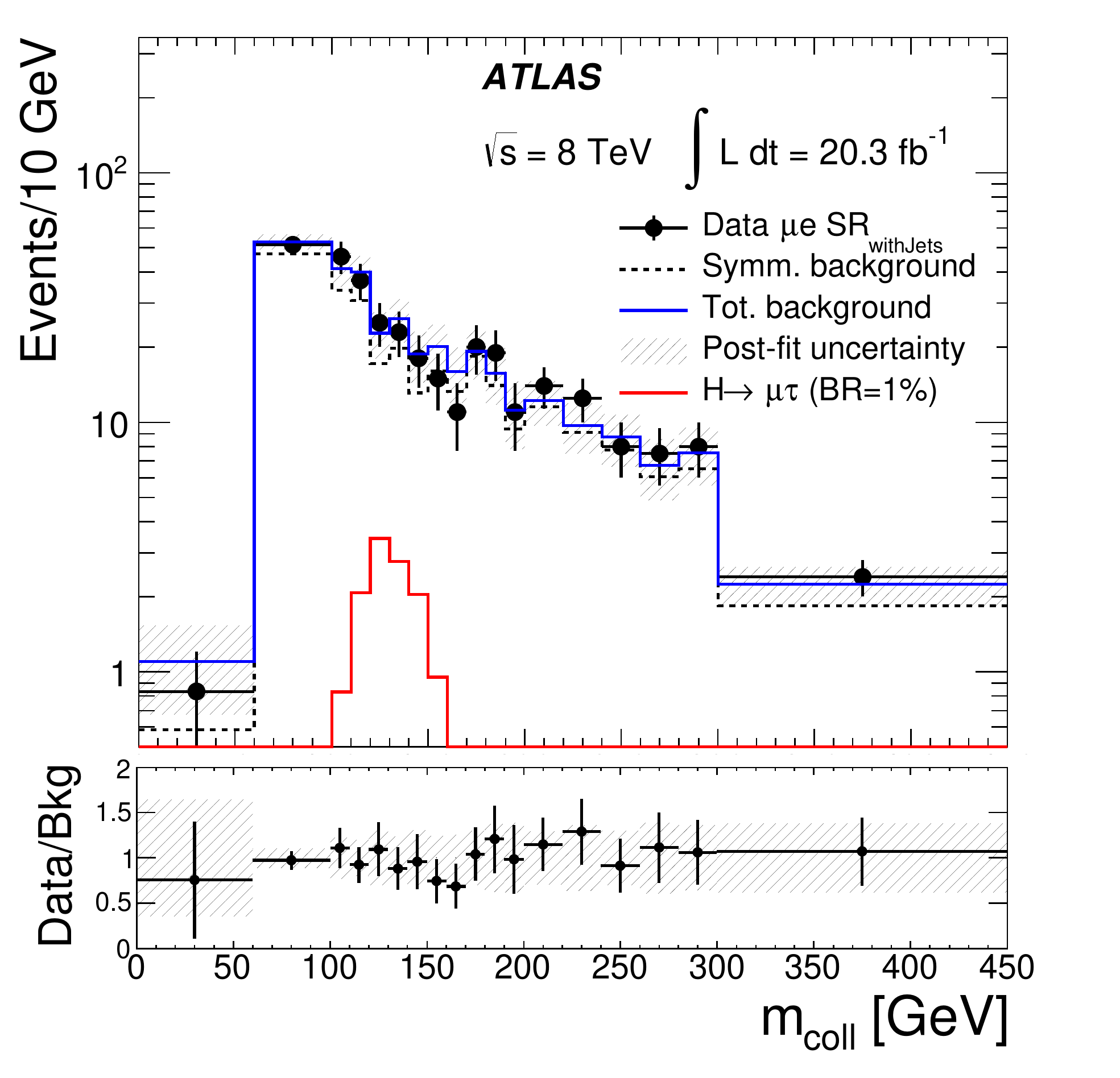}
\caption{Collinear mass distributions in the $\tlep{}$ channel: background
  estimate compared to the events observed in the data in the
  \SRnoJets{}~(top) and \SRwithJets{}~(bottom). Left:
  \ems{}~channel. Right: \mes{}~channel.  In these plots, events from
  the three \fptOne{} bins are combined, although the fit parameters
  are different in each \fptOne{} bin. The signal expected for a
  $\text{Br}(\htm{})=1\%$ is shown in the \mes{} channel.
}
\label{fig:yieldsData_combined}
\end{center}
\end{figure}

Figure~\ref{fig:yieldsData_combined} compares the observed data to
the yields expected from the symmetry-based statistical model. The
comparison, combining the different \ptOne{} bins, shows the symmetric
component of the background ($b_{ij}$) as a dashed line, and the
total background estimation including the contribution from events
containing misidentified and non-prompt leptons as a full line. As can be seen,
the background estimation is in good agreement with the data over the
full mass range.  Table~\ref{tab:dataYields} summarises the fit results
in the data in $\mathrm{SR_{noJets}}$~and
$\mathrm{SR_{withJets}}$: the fitted \fptOne{}~scale
factors, the symmetric background component
($\sum\limits_{i}^{N_{\mcol}}b_{ij}$) in each \ptOne{}~bin,
and the non-prompt estimate in the \mes{}~and the \ems{} channels.
The excellent level of agreement between the fitted number of events and the
observed number is due to the many unconstrained parameters
in the fit.
%
%
\begin{table}[htb]
  \caption{A summary of the fit results in the $\tlep{}$ channel.
    The values of the fit parameters \fptOne{}, which account for the
    ratio of the \ems{} and \mes{} event reconstruction efficiencies
    described in Section~\ref{sec:ll_bkg}, are obtained from a
    background-only fit, and reported for each signal region and for
    each \ptOne{} bin.
    The expected and observed yields correspond to the number of
    events used in the fit, representing the 0--300~\GeV{} \mcol{}
    range shown in Figure~\ref{fig:yieldsData_combined}.
    The quoted uncertainties in the expected yields represent the
    statistical (first) and systematic (second) uncertainties,
    respectively.
    The post-fit values of systematic uncertainties are 
    provided for the background predictions. 
    The signal predictions are given for
    $\text{Br}(\hte{})=1\%$ in the \ems{} sample and for
    $\text{Br}(\htm{})=1\%$ in the \mes{} sample.
  }
\centering
\label{tab:dataYields}
\begin{tabular}{lcc r r@{$\,\pm\,$}r@{$\,\pm$}r r}
\toprule
\SRnoJets{} \\
\cmidrule(lr){1-2}
\ptOne{}~bin [\GeV{}]   & \fptOne{}                           &        & LFV Signal, Br=1\%      & \multicolumn{3}{c}{Total Backg.} & Observed \\
\midrule
\multirow{2}{*}{12--20} & \multirow{2}{*}{$1.11 \pm 0.06$}    & $\ems$ & $ 14.9 \pm 0.4 \pm 2.7$ & 1219 & 24 & 27  & $1212$ \\
                        &                                     & $\mes$ & $ 10.7 \pm 0.4 \pm 2.3$ & 1033 & 25 & 20  & $1035$ \\
\cmidrule(lr){1-2}
\multirow{2}{*}{20--30} & \multirow{2}{*}{$1.07 \pm 0.08$}    & $\ems$ & $ 15.1 \pm 0.4 \pm 2.7$ &  998 & 22 & 25  & $ 995$ \\
                        &                                     & $\mes$ & $ 12.4 \pm 0.4 \pm 2.2$ &  950 & 23 & 21  & $ 950$ \\
\cmidrule(lr){1-2}
\multirow{2}{*}{$\geq 30$} & \multirow{2}{*}{$1.01 \pm 0.07$} & $\ems$ & $ 12.5 \pm 0.4 \pm 2.2$ &  455 & 17 & 16  & $ 452$ \\
                           &                                  & $\mes$ & $ 11.4 \pm 0.4 \pm 2.0$ &  458 & 16 & 14  & $ 457$ \\
\midrule
\SRwithJets{} \\
\cmidrule(lr){1-2}
\ptOne{}~bin [\GeV{}]   & \fptOne{}                           &        & LFV Signal, Br=1\%     & \multicolumn{3}{c}{Total Backg.}   & Observed \\
\midrule
\multirow{2}{*}{12--20} & \multirow{2}{*}{$1.07 \pm 0.10$}    & $\ems$ & $ 5.9 \pm 0.3 \pm 1.1$ &  222 & 10 & 11 & $220$ \\
                        &                                     & $\mes$ & $ 3.9 \pm 0.2 \pm 0.9$ &  181 & 10 &  9 & $182$ \\
\cmidrule(lr){1-2}
\multirow{2}{*}{20--30} & \multirow{2}{*}{$1.24 \pm 0.16$}    & $\ems$ & $ 5.4 \pm 0.2 \pm 1.1$ &  187 &  9 & 11 & $187$ \\
                        &                                     & $\mes$ & $ 4.5 \pm 0.2 \pm 0.9$ &  161 &  9 &  9 & $161$ \\
\cmidrule(lr){1-2}
\multirow{2}{*}{$\geq 30$} & \multirow{2}{*}{$1.13 \pm 0.10$} & $\ems$ & $ 5.5 \pm 0.2 \pm 1.0$ &  251 & 11 & 12 & $250$ \\
                           &                                  & $\mes$ & $ 4.9 \pm 0.2 \pm 0.9$ &  229 & 11 & 11 & $229$ \\
\bottomrule
\end{tabular}
\end{table}
The expected and observed 95$\%$ CL upper limits on branching
ratios as well as their best fit values are calculated using the
statistical model described in Section~\ref{sec:leplepStat}.
Table~\ref{tab:combinedLimits} presents a summary of results for the
individual categories and their combination can be found in
Table~\ref{tab:combinedLimits} for both the \hte{} and \htm{}
hypotheses.

%% file: paper_results.tex
\section{Combined results of the search for LFV $H\to e\tau/\mu\tau$ decays}
\label{sec:results}


\input combined_results

\FloatBarrier


%% file: combined_results.tex

\begin{figure}[htb]
\begin{center}
\subfloat[]{
\includegraphics[width=0.45\textwidth]{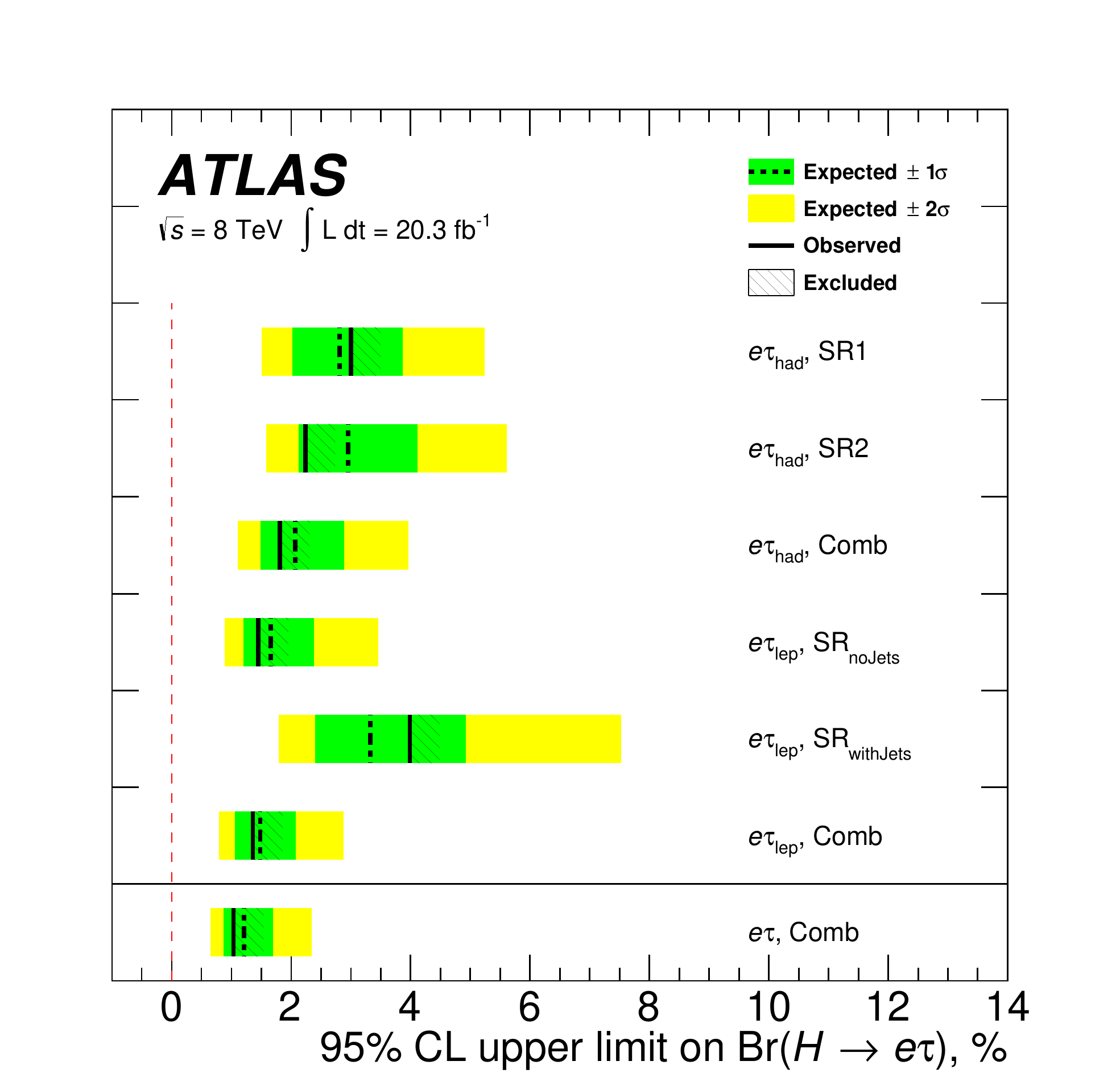}
}
\subfloat[]{
\includegraphics[width=0.45\textwidth]{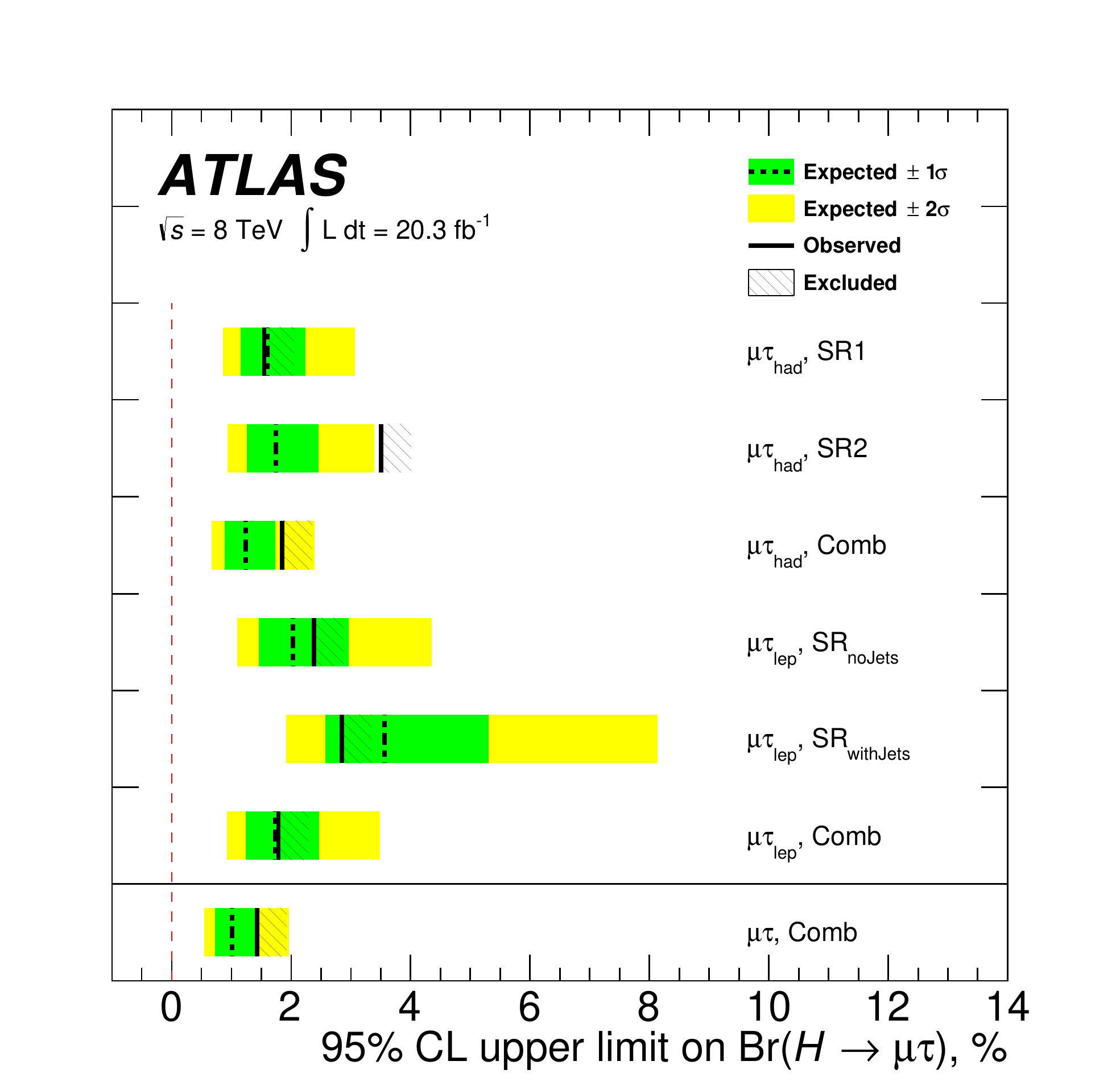}
}
\caption{Upper limits on LFV decays of the Higgs boson in the \hte{}
  hypothesis (left) and \htm{} hypothesis (right).
  The limits are computed under the assumption that either
  Br($H\to{}\mu{}\tau{}$)=0 or Br($H\to{}e\tau{}$)=0.
The $\mu\thad$ channel is from Ref.~\cite{Aad:2015gha}.}
\label{fig:hlfv_summary}
\end{center}
\end{figure}

The results of the individual searches for the LFV $H\to e\tau$ and
$H\to\mu\tau$ decays in the $\thad$ (including the result from
Ref.~\cite{Aad:2015gha}) and $\tlep$ channels presented in
Sections~\ref{sec:lephad_results} and~\ref{sec:leplep_results} are
statistically combined. The two channels use different background 
estimation techniques, leading to uncorrelated systematic uncertainties 
in the background predictions. The systematic uncertainties for the
LFV signal are treated as 100$\%$ correlated between the two channels.  
Table~\ref{tab:combinedLimits} presents a summary of results for the
expected and observed $95\%$ CL upper limits and the best fit values
for the branching ratios for the individual categories and their
combination. There is no indication of a signal
in the search for the LFV $H\to e\tau$ decays. The combined 
observed, and the median expected, 95$\%$ CL upper limits on Br($H\to e\tau$) 
for a Higgs boson with $m_{H} = 125$~GeV are 1.04$\%$ and 
$1.21^{+0.49}_{-0.34}$$\%$, respectively. A small $\sim$1$\sigma$ excess 
of data over the predicted background is observed in the search for the 
LFV $H\to\mu\tau$ decays. It is mostly driven by a 1.3$\sigma$ excess 
in the earlier search in the $\mu\thad$ channel~\cite{Aad:2015gha}. This 
corresponds to a best fit value for the branching ratio of 
Br($H\to\mu\tau$) = ($0.53 \pm 0.51$)$\%$. In the absence of any
significant signal, an upper limit on the LFV branching ratio
Br($H\to\mu\tau$) for a Higgs boson with $m_\mathrm{H} = 125$~GeV is
set. The corresponding observed, and the median expected, 95$\%$ CL upper 
limits are 1.43$\%$ and $1.01^{+0.40}_{-0.29}$$\%$, respectively.
The upper limits on the LFV decays of the Higgs boson are summarised
in Figure~\ref{fig:hlfv_summary}.

%

{
\renewcommand{\arraystretch}{1.5}
\begin{table}[htb]
  \begin{center}
    \caption{Results of the search for the LFV $H\to e\tau$ and $H\to\mu\tau$ decays.
      The limits are computed under the assumption that either
      Br($H\to{}\mu{}\tau{}$)=0 or Br($H\to{}e\tau{}$)=0.
      The expected and observed $95\%$ confidence level (CL) upper limits and the best
      fit values for the branching ratios for the individual categories and their combination.
      The $\mu\thad$ channel is from Ref.~\cite{Aad:2015gha}.
    \label{tab:combinedLimits}}
    \begin{tabular}{lccc r@{.}l@{}l}
       \toprule
Channel        &  Category  &  Expected limit [$\%$] & Observed limit [$\%$] & \multicolumn{3}{c}{Best fit Br [\%]} \\ 
               \midrule
               &  SR1       & $2.81^{+1.06}_{-0.79}$   & 3.0      & $ 0$&33&$^{+1.48}_{-1.59}$ \\
$H\to e\thad$  &  SR2       & $2.95^{+1.16}_{-0.82}$   & 2.24     & $-1$&33&$^{+1.56}_{-1.80}$ \\
               &  Combined  & $2.07^{+0.82}_{-0.58}$   & 1.81     & $-0$&47&$^{+1.08}_{-1.18}$ \\ 
\cmidrule(lr){2-7}
               & \SRnoJets   & $1.66^{+0.72}_{-0.46}$  & 1.45     & $-0$&45&$^{+0.89}_{-0.97}$ \\
$H\to e\tlep$  & \SRwithJets & $3.33^{+1.60}_{-0.93}$  & 3.99     & $ 0$&74&$^{+1.59}_{-1.62}$ \\
               &  Combined   & $1.48^{+0.60}_{-0.42}$  & 1.36     & $-0$&26&$^{+0.79}_{-0.82}$ \\ 
\cmidrule(lr){2-7}  
$H\to e\tau$   &  Combined   & $1.21^{+0.49}_{-0.34}$   & 1.04    & $-0$&34&$^{+0.64}_{-0.66}$ \\  
\midrule
               &  SR1        & $1.60^{+0.64}_{-0.45}$   & 1.55    & $-0$&07&$^{+0.81}_{-0.86}$ \\
$H\to\mu\thad$ &  SR2        & $1.75^{+0.71}_{-0.49}$   & 3.51    & $ 1$&94&$^{+0.92}_{-0.89}$ \\
               &  Combined   & $1.24^{+0.50}_{-0.35}$   & 1.85    & $ 0$&77&$^{+0.62}_{-0.62}$ \\
\cmidrule(lr){2-7} 
               & \SRnoJets   & $2.03^{+0.93}_{-0.57}$  & 2.38     & $ 0$&31&$^{+1.06}_{-0.99}$ \\
$H\to\mu\tlep$ & \SRwithJets & $3.57^{+1.74}_{-1.00}$  & 2.85     & $-1$&03&$^{+1.66}_{-1.82}$ \\
               &  Combined   & $1.73^{+0.74}_{-0.49}$  & 1.79     & $ 0$&03&$^{+0.88}_{-0.86}$ \\ 
\cmidrule(lr){2-7}
$H\to \mu\tau$  &  Combined  & $1.01^{+0.40}_{-0.29}$   & 1.43    & $ 0$&53&$^{+0.51}_{-0.51}$ \\   
\bottomrule
    \end{tabular}
  \end{center}
\end{table}
}


%% file: Z_selection.tex
\section{Search for \Ztaumu{} using the $\thad$ channel}

The search for \Ztaumu{} events is based on $\mu\thad$ final state and utilises the same strategy as the $H\to\mu\tau$ analysis documented  in Ref.~\cite{Aad:2015gha}, and applied to the  $H\to e \thad$ search described above. The final state is characterised by the presence of an energetic muon and a $\thad$ of opposite charge and the presence of moderate $\MET{},$ aligned with the $\thad$ direction. The typical transverse momenta of the muon and of the $\thad$ are somewhat softer than those expected in Higgs boson LFV decay, due to the lower mass of the $Z$ boson. 
The main backgrounds are the same as those observed in  $H\to\mu\thad$ analyses, namely: $Z\to\tau\tau$, $W$+jets, multi-jet, $H\to\tau\tau$, diboson and top backgrounds. The \Mmutau{} variable is used to extract the signal using the same fit procedure and estimation of systematic uncertainties as for the $H\to \mu \thad$ search. The corresponding Higgs boson LFV contribution is assumed to be negligible. 

The \Ztaumu{} analysis differs from the $H\to\mu\thad$ one as follows:
\begin{itemize}
\item The signal and control regions are defined in the same way as in the $H\to\mu\thad$ analysis, but the cut values are lowered to match the kinematics of $Z$ boson decay products. The exact definition is given in Table~\ref{tab:Z_cuts}.
\item The LFV $H\to\mu\thad$ signal sample is replaced with a LFV \Ztaumu{} signal sample.
\item The shape correction for $W$+jets in SR1 is obtained from the $\Mmutau>110$ \GeV{} sideband in SR1.
\item Due to larger $W$+jets contribution in SR1 and SR2,
  the shape corrections for the $W$+jets samples are calculated using a three-dimensional binning
  scheme in $\pt(\thad)$, $|\eta(\mu)-\eta(\thad)|$ and
  $N_{\mathrm{jet}}$.
\item The $W$+jets 
 extrapolation  uncertainty, which accounts for
  the difference between the $W$+jets ALPGEN PYTHIA and HERWIG samples,
  is also included as a shape uncertainty.
\end{itemize}

\begin{figure}
\centering
\includegraphics[width=0.49\textwidth]{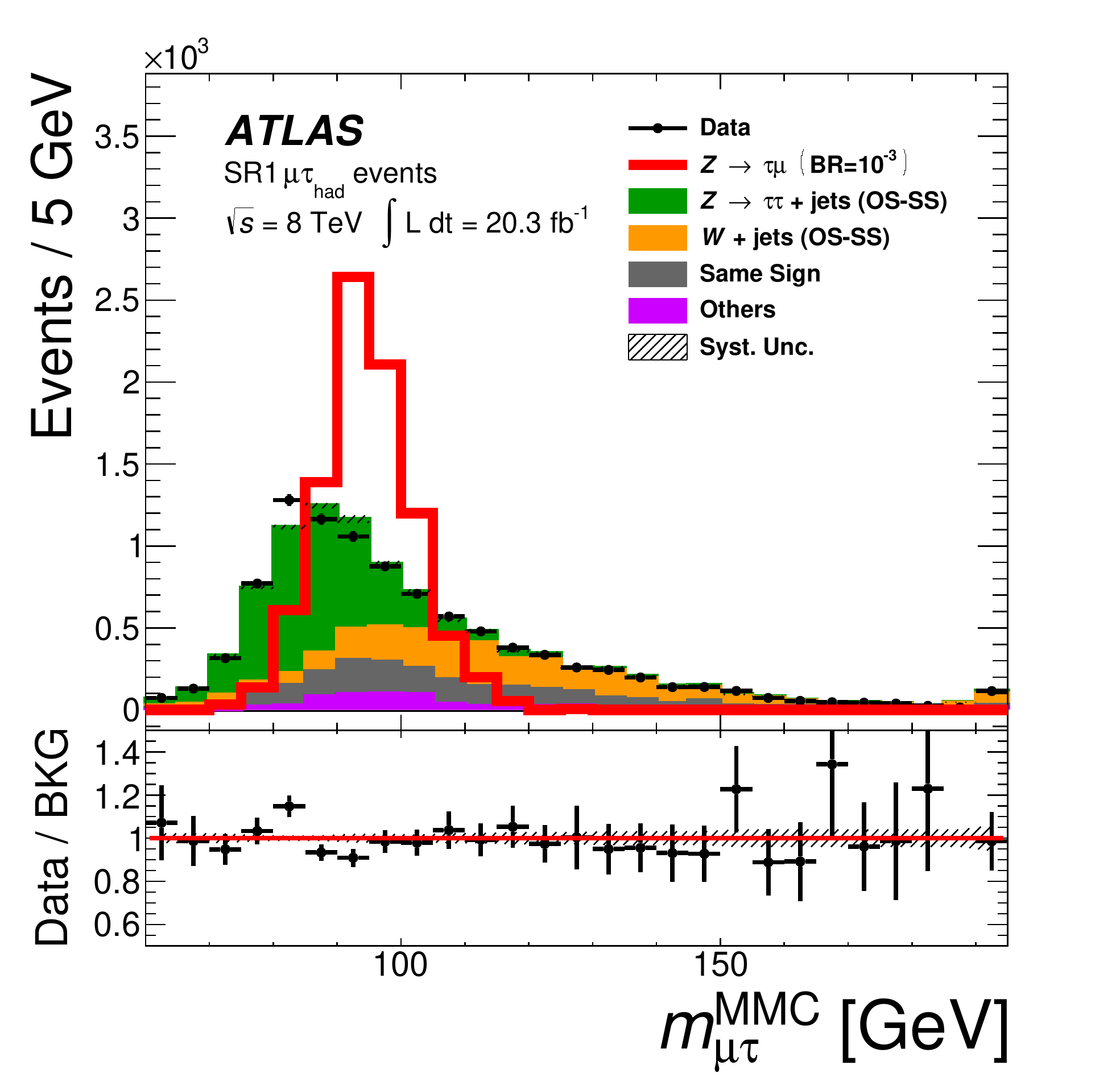}
\includegraphics[width=0.49\textwidth]{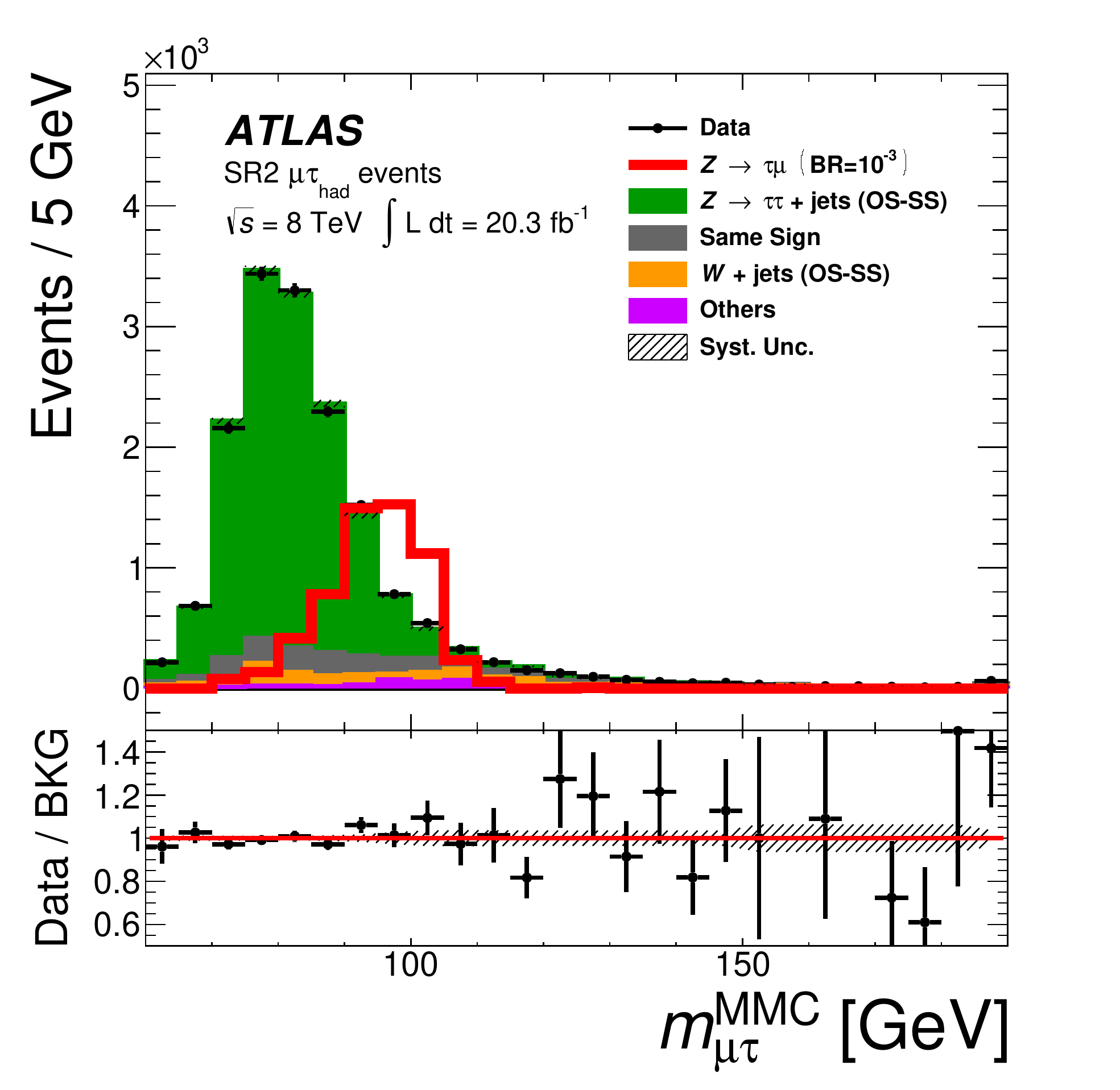}
\caption{Distributions of the mass reconstructed by the Missing Mass
  Calculator, $\Mmutau$, in \Ztaumu{} SR1 (left) and SR2 (right). 
  The background distributions are determined in a global fit.
  The signal distributions are scaled to a branching ratio of
  Br($Z\to\mu\tau$) = $10^{-3}$ to make them visible. The 
  bottom panel of each subfigure shows the ratio 
  of the observed data to the estimated background. 
  The hatched band for the ratio illustrates post-fit systematic 
  uncertainties in the background prediction. The statistical
  uncertainties for data and background predictions are added 
  in quadrature for the ratios. The last bin of the distribution contains events with $\Mmutau>200$ \GeV{}.}
\label{fig:ZLFV}
\end{figure}

The numbers of observed events and background in each of the regions
are given in Table~\ref{tab:Z_yields}.  The efficiencies for simulated
\Ztaumu{} signal events to pass the SR1 and SR2 selections are 1.2\%
and 0.8\%, respectively.  Figure~\ref{fig:ZLFV} shows the \Mmutau{}
distribution for data and predicted background in each of the signal
regions.
The discrepancy observed in the \Mmutau{} range 80--100~\GeV{} of SR1
was studied carefully. All the other SR1 distributions, including
lepton momenta, transverse masses, and missing transverse momentum,
are in excellent agreement with the predictions, and the background
shapes are constrained in the control regions as well as in SR2.
This discrepancy is hence attributed to a statistical fluctuation.
\begin{table}[htb]
  \begin{center}
    \caption{Summary of the \Ztauhmu{} event selection criteria used to define
      the signal and control regions (see text).                         
    \label{tab:Z_cuts}}
    \begin{tabular}{cllll}
      \toprule
Cut                                     & SR1                                             & SR2                  & WCR        & TCR \\  \midrule
$\pt(\mu)$                         & $>$30~\GeV{}                                & $>$30~\GeV{}     & $>$30~\GeV{}     & $>$30~\GeV{} \\ 
$\pt(\thad)$                       & $>$30~\GeV{}                                & $>$30~\GeV{}     & $>$30~\GeV{}     & $>$30~\GeV{}  \\
$|\eta(\mu)-\eta(\thad)|$  & $<$2                                          & $<$2                & $<$2             & $<$2        \\
$\mT^{\mu,\met}$               & $>$30~\GeV{}  and $<$75~\GeV{}     & $<$30~\GeV{}     & $>$60~\GeV{}    & --  \\
$\mT^{\thad,\met}$             & $<$20~\GeV{}                                & $<$45~\GeV{}     & $>$40~\GeV{}    & --   \\
$N_{\mathrm{jet}}$            &  --  &  --  &  --  & $>$1    \\
$N_{b-\mathrm{jet}}$          &  0  &  0  &  0  & $>$0    \\  \bottomrule
    \end{tabular}
  \end{center}
\end{table}

%

\begin{table}[htb]
  \begin{center}
    \caption{Data yields, signal and post-fit OS--SS background
      predictions (see Eq.~(\ref{eq:bckg_formula})) for the 
       \Ztauhmu{} 80~\GeV{}$<\Mmutau<$115~\GeV{} region. The signal predictions are 
      given assuming Br(\Ztaumu) = $10^{-5}$. 
      The background predictions are obtained from the combined fit
      to SR1, SR2, WCR and TCR.   To calculate these quantities
      for SR1 and SR2, the signal strengths are decorrelated in the
      signal regions and set to zero in the control regions.
      The post-fit values of systematic uncertainties are 
      provided for the background predictions. 
      For the total background, all correlations between various sources 
      of systematic uncertainties and backgrounds are taken into account.
      The quoted uncertainties represent the statistical (first) and
      systematic (second) uncertainties, respectively.                           
    \label{tab:Z_yields}}
    \begin{tabular}{l r@{}l@{}c@{}r@{}l@{}c@{}r@{}l r@{}l@{}c@{}r@{}l@{}c@{}r@{}l}
      \toprule
                              & \multicolumn{8}{c}{SR1}                 & \multicolumn{8}{c}{SR2}                \\  \midrule
      Signal                  &    86&   &$\pm$&   2&   &$\pm$&  22&    &   56&   &$\pm$&   2&   &$\pm$&  18&    \\  \midrule
      $Z\to\tau\tau$          &  3260&   &$\pm$&  30&   &$\pm$&  60&    & 7060&   &$\pm$&  40&   &$\pm$& 150&    \\
      $W$+jets                &  1350&   &$\pm$&  70&   &$\pm$& 110&    &  590&   &$\pm$&  50&   &$\pm$&  70&    \\
      Same--Sign events       &  1110&   &$\pm$&  40&   &$\pm$& 100&    &  930&   &$\pm$&  30&   &$\pm$&  90&    \\
      $VV+Z\to\mu\mu$         &   410&   &$\pm$&  60&   &$\pm$&  50&    &  240&   &$\pm$&  60&   &$\pm$&  60&    \\
      $H\to\tau\tau$          &    25&.1 &$\pm$&   0&.5 &$\pm$&   3&.0  &   41&   &$\pm$&   1& &$\pm$&   5&  \\
      Top                     &    22&   &$\pm$&   4&   &$\pm$&   4&    &   15&   &$\pm$&   4&   &$\pm$&   4&  \\ \midrule
      Total background        &  6170&   &$\pm$& 100&   &$\pm$& 100&    & 8880&   &$\pm$& 100&   &$\pm$& 140&    \\ \midrule
      Data                    &  6134&   &     &    &   &     &    &    & 8982&   &     &    &   &     &    &    \\ \bottomrule
    \end{tabular}
  \end{center}
\end{table}
%
%
No excess of data is observed and the CL$_s$ limit-setting
technique is used to calculate the observed and expected limits on the branching ratio for
\Ztaumu$\,$ decays. 
The observed 95 \% CL limit on Br($Z\to\mu\tau$) is $1.7 \times 10^{-5}$, which is
lower than the expected upper limit of
Br($Z\to\mu\tau$)$ = 2.6 \times 10^{-5}$, but still within the $2 \sigma$ band.
This corresponds to a best fit value for the branching ratio
Br($Z\to\mu\tau$)$ = -1.6_{-1.4}^{+1.3} \times 10^{-5}$.  The results for the different
signal regions are summarised in Table~\ref{tab:resultsZtaumu}.
%
%


\begin{table}[htb]
  \small
  \begin{center}     
    \caption{The expected and observed 95\% CL exclusion limits as
      well as the best fit values for the branching ratio of
      $\text{Br}(Z\to\mu\tau)[10^{-5}]\,$ are shown
      for  SR1, SR2 and the combined fit. To calculate these quantities
      for SR1 and SR2, the signal strengths are decorrelated in the
      signal regions and set to zero in the control regions.}
    \label{tab:resultsZtaumu}
    \begin{tabular}{l r@{.}l@{}l r@{.}l@{}l r@{.}l@{}l}
      \toprule   
      $\text{Br}(Z\to\mu\tau)[10^{-5}]$ & \multicolumn{3}{c}{SR1} & \multicolumn{3}{c}{SR2} & \multicolumn{3}{c}{Combined} \\
      \midrule 
      Expected limit    &$ 2$&6&$_{-0.7}^{+1.1}$        &$6$&4&$^{-1.8}_{+2.8}$   &$ 2$&6&$_{-0.7}^{+1.1}$ \\
      Observed limit    &$ 1$&5&                        &$7$&9&                   &$ 1$&7&                 \\
      Best  fit         &$-2$&1&$^{+1.2}_{-1.3}$        &$2$&6&$^{+2.9}_{-2.6}$   &$-1$&6&$_{-1.4}^{+1.3}$ \\
      \bottomrule
    \end{tabular}
  \end{center} 
\end{table}

%



%% file: paper_summary.tex
\section{Summary}
\label{sec:summary}

Searches for lepton-flavour-violating decays of the $Z$ and
Higgs bosons are performed using a data sample of proton--proton 
collisions recorded by the ATLAS detector at the LHC corresponding to an
integrated luminosity of 20.3~\ifb\ at $\sqrt{s}=8$~\TeV{}. Three LFV decays 
are considered: $H\to e\tau$, $H\to\mu\tau$, and $Z\to\mu\tau$. The search 
for the Higgs boson decays is performed in the final states where the 
$\tau$-lepton decays either to hadrons or to leptons (electron or muon). 
The search for the $Z$ boson decays is performed in the final 
state with the $\tau$-lepton decaying into hadrons. No significant excess 
is observed, and upper limits on the LFV branching ratios are set. The 
observed and the median expected 95$\%$ CL upper limits on Br($H\to e\tau$) 
are \LimObsHe$\%$ and \LimExpHe$\%$, respectively. This
direct search for the $H\to e\tau$ decays places significantly
more stringent constraints on Br($H\to e\tau$) than earlier
indirect estimates. In the search for the $H\to\mu\tau$ decays, the 
observed and the median expected 95$\%$ CL upper limits on Br($H\to\mu\tau$) 
are \LimObsHmu$\%$ and \LimExpHmu$\%$, respectively. A small deficit of data compared
to the predicted background is observed in the search for the LFV
$Z\to\mu\tau$ decays. The observed and the median expected 95$\%$ CL upper 
limits on Br($Z\to\mu\tau$) are $\LimObsZmu$ and $\LimExpZmu$, respectively.



%% file: Acknowledgements.tex

We thank CERN for the very successful operation of the LHC, as well as the
support staff from our institutions without whom ATLAS could not be
operated efficiently.

We thank Avital Dery and Aielet Efrati for their significant contribution and dedication to this study.

We acknowledge the support of ANPCyT, Argentina; YerPhI, Armenia; ARC, Australia; BMWFW and FWF, Austria; ANAS, Azerbaijan; SSTC, Belarus; CNPq and FAPESP, Brazil; NSERC, NRC and CFI, Canada; CERN; CONICYT, Chile; CAS, MOST and NSFC, China; COLCIENCIAS, Colombia; MSMT CR, MPO CR and VSC CR, Czech Republic; DNRF and DNSRC, Denmark; IN2P3-CNRS, CEA-DSM/IRFU, France; GNSF, Georgia; BMBF, HGF, and MPG, Germany; GSRT, Greece; RGC, Hong Kong SAR, China; ISF, I-CORE and Benoziyo Center, Israel; INFN, Italy; MEXT and JSPS, Japan; CNRST, Morocco; FOM and NWO, Netherlands; RCN, Norway; MNiSW and NCN, Poland; FCT, Portugal; MNE/IFA, Romania; MES of Russia and NRC KI, Russian Federation; JINR; MESTD, Serbia; MSSR, Slovakia; ARRS and MIZ\v{S}, Slovenia; DST/NRF, South Africa; MINECO, Spain; SRC and Wallenberg Foundation, Sweden; SERI, SNSF and Cantons of Bern and Geneva, Switzerland; MOST, Taiwan; TAEK, Turkey; STFC, United Kingdom; DOE and NSF, United States of America. In addition, individual groups and members have received support from BCKDF, the Canada Council, CANARIE, CRC, Compute Canada, FQRNT, and the Ontario Innovation Trust, Canada; EPLANET, ERC, FP7, Horizon 2020 and Marie Sk{\l}odowska-Curie Actions, European Union; Investissements d'Avenir Labex and Idex, ANR, R{\'e}gion Auvergne and Fondation Partager le Savoir, France; DFG and AvH Foundation, Germany; Herakleitos, Thales and Aristeia programmes co-financed by EU-ESF and the Greek NSRF; BSF, GIF and Minerva, Israel; BRF, Norway; Generalitat de Catalunya, Generalitat Valenciana, Spain; the Royal Society and Leverhulme Trust, United Kingdom.

The crucial computing support from all WLCG partners is acknowledged
gratefully, in particular from CERN, the ATLAS Tier-1 facilities at
TRIUMF (Canada), NDGF (Denmark, Norway, Sweden), CC-IN2P3 (France),
KIT/GridKA (Germany), INFN-CNAF (Italy), NL-T1 (Netherlands), PIC
(Spain), ASGC (Taiwan), RAL (UK) and BNL (USA), the Tier-2 facilities
worldwide and large non-WLCG resource providers. Major contributors of
computing resources are listed in Ref.~\cite{ATL-GEN-PUB-2016-002}.

%% file: atlas_authlist.tex
\begin{flushleft}
{\Large The ATLAS Collaboration}

\bigskip

G.~Aad$^{\rm 87}$,
B.~Abbott$^{\rm 114}$,
J.~Abdallah$^{\rm 65}$,
O.~Abdinov$^{\rm 12}$,
B.~Abeloos$^{\rm 118}$,
R.~Aben$^{\rm 108}$,
M.~Abolins$^{\rm 92}$,
O.S.~AbouZeid$^{\rm 138}$,
N.L.~Abraham$^{\rm 150}$,
H.~Abramowicz$^{\rm 154}$
% ...
\bigskip
\\
% ...
$^{12}$ Institute of Physics, Azerbaijan Academy of Sciences, Baku, Azerbaijan\\
% ...
$^{65}$ University of Iowa, Iowa City IA, United States of America\\
% ...
$^{87}$ CPPM, Aix-Marseille Universit{\'e} and CNRS/IN2P3, Marseille, France\\
% ...
$^{92}$ Department of Physics and Astronomy, Michigan State University, East Lansing MI, United States of America\\
% ...
$^{108}$ Nikhef National Institute for Subatomic Physics and University of Amsterdam, Amsterdam, Netherlands\\
% ...
$^{114}$ Homer L. Dodge Department of Physics and Astronomy, University of Oklahoma, Norman OK, United States of America\\
% ...
$^{118}$ LAL, Univ. Paris-Sud, CNRS/IN2P3, Universit{\'e} Paris-Saclay, Orsay, France\\
% ...
$^{138}$ Santa Cruz Institute for Particle Physics, University of California Santa Cruz, Santa Cruz CA, United States of America\\
% ...
$^{150}$ Department of Physics and Astronomy, University of Sussex, Brighton, United Kingdom\\
% ...
$^{154}$ Raymond and Beverly Sackler School of Physics and Astronomy, Tel Aviv University, Tel Aviv, Israel\\
% ...
\end{flushleft}
